\documentstyle[prb,eqsecnum,aps,epsf]{revtex}
\begin{document}
\draft


\title{
Resonant tunneling and the multichannel Kondo problem: 
the quantum Brownian motion description}

\author{Hangmo Yi}

\address{Department of Physics and Astronomy,
 University of Kentucky \\
 Lexington, KY 40506-0055}

\date{\today}

\maketitle

\begin{abstract}
We study mesoscopic resonant tunneling as well 
as multichannel Kondo problems by mapping 
them to a first-quantized quantum mechanical model of a particle 
moving in a multi-dimensional periodic potential 
with Ohmic dissipation.  
From a renormalization group analysis, 
we obtain phase diagrams of the quantum 
Brownian motion model with various lattice symmetries.  
For a symmorphic lattice, there are 
two phases at $T=0$: a localized phase in 
which the particle is trapped in a potential 
minimum, and a free phase in which 
the particle is unaffected by the periodic potential.  
For a non-symmorphic lattice, however, there 
may be an additional intermediate phase in which the 
particle is neither localized nor completely free.  
The fixed 
point governing the intermediate phase is shown 
to be identical to the well-known multichannel 
Kondo fixed point in the Toulouse limit 
as well as the resonance fixed point of 
a quantum dot model and a double-barrier Luttinger 
liquid model.  The mapping allows us to 
compute the fixed-point mobility $\mu^*$ of the quantum 
Brownian motion model exactly, using known 
conformal-field-theory results of the Kondo problem.  
From the mobility, we find that the peak value of the 
conductance resonance of a spin-1/2 quantum 
dot problem is given by $e^2/2h$.  The scaling 
form of the resonance line shape is predicted.  
\end{abstract}
\pacs{PACS numbers: 05.40.+j, 05.30.$-$d, 72.15.Qm, 73.40.Gk}

\tableofcontents

\section{Introduction}
\label{sec:intro}

Electronic transport in mesoscopic systems has 
drawn much attention in the last decade 
due to recent development of 
fabrication techniques of ultra small structures of 
high quality.  In the nanometer scale systems, 
electron-electron interactions play more 
important roles, as the number of effective spatial 
dimensions $d$ is lowered.  Most remarkable among many 
examples are fractional quantum Hall effect 
and Wigner crystallization at $d=2$, \cite{dasSarma97book} 
Luttinger liquid behavior 
at $d=1$, \cite{tarucha95,milliken96,chang96} and Coulomb 
blockade at $d=0$. \cite{grabert92book}
On the theoretical side, the use of traditional 
perturbative approaches were proven less useful 
due to the strong correlations, while 
nonperturbative techniques such as renormalization 
group, conformal field theory, and the Bethe 
ansatz have seen some success.  

Studying resonant tunneling in a quantum dot 
coupled to leads via two identical 
quantum point contacts (QPC), 
Furusaki and Matveev \cite{furusaki95} showed that 
at $T=0$, the on-resonance conductance 
of a spin-1/2 system 
takes a universal value less than the perfect 
conductance $e^2/h$. \cite{perfectConductance}  
They argued that the on-resonance properties are 
controlled by the highly nontrivial 
non-Fermi-liquid fixed point of a four-channel 
Kondo problem.  The deep connection between 
the problem of resonant tunneling 
in a point-contact-coupled quantum 
dot structure and the Kondo problem has been 
known for some time. \cite{glazman90,matveev91} 
A recent experiment \cite{berman99} has confirmed 
the theoretical predictions \cite{furusaki95a} 
of the capacitance line shape which 
was based on this connection.  

Later, Yi and Kane \cite{yi98} 
showed that there is a general and 
direct one-to-one mapping between 
the resonant tunneling problem and 
the Kondo problem in a special limit.  
They argued that this limit is a multichannel 
generalization of the 
well-known Toulouse limit. \cite{toulouse70}  Using 
the conformal-field-theory results on the Kondo 
fixed point, they computed the 
universal conductance on resonance exactly.  

They have also found that the quantum-field-theoretical 
formalism of both problems can be reduced to a 
first-quantized quantum 
mechanical formalism of a {\em fictitious} particle 
that moves in a periodic potential in a dissipative 
medium. \cite{caldeira83,schmid83,fisher85}  
Technically, this single-particle 
quantum Brownian motion (QBM) model 
is obtained by integrating out all degrees of freedom 
in the original quantum field theory except 
only a small finite number of them.  These remaining 
degrees of freedom become the ``spatial coordinates'' of 
the Brownian particle and the integrated ones 
serve as the source of effective dissipation.  

The symmetry of the periodic potential in 
the QBM model has been found to be 
crucial in identifying the zero-temperature phase 
diagram, while a particular phase is determined 
by the potential period and the dissipation strength.  
In particular, the on-resonance quantum dot problem 
were related to potentials with non-symmorphic 
lattice symmetries.  A similar 
QBM model has been used in relatively 
recent studies in the context of resonance tunneling 
through an impurity state 
in a Luttinger liquid, \cite{kane92,kane92a} 
which apply to such experimental situations 
as in quantum wires \cite{tarucha95,yacoby96} and 
quantum Hall edge systems. \cite{milliken96}  
More applications of the QBM descriptions 
may also be found in connection with tunneling between 
quantum Hall edge states, both in 
fractional \cite{moon93} and 
integer \cite{yi96} quantum Hall regime.  
A similar approach has also been used to 
study resonant point-contact tunneling 
between Luttinger-liquid leads, \cite{nayak99} 
but the QBM model in that work 
was formulated with a $\pi$-flux through 
each plaquette.  

In this paper, we extend the investigations in 
Ref.\ \ref{ref:yi98} and present a more 
general description of the 
QBM model as applied to (i) resonant 
tunneling in a quantum dot 
coupled to an arbitrary number of leads, 
(ii) multichannel Kondo problem, 
and (iii) resonant tunneling of Luttinger liquid through 
a double-barrier.  As will be shown below, 
the QBM model possesses very rich phase diagrams.  
There are three zero-temperature phases: (i) the ``free'' 
phase in which the Brownian particle completely ignores 
the periodic potential, (ii) the ``localized'' 
phase in which the particle is trapped in a 
minimum of the periodic potential, (iii) and the most exotic 
of them all, the ``intermediate'' phase in which 
the particle is neither localized nor completely free.  
Each phase is characterized by 
a ``mobility'' $\mu$ --- the ratio of the 
average velocity of the Brownian particle 
to the driving force in the linear response regime 
--- which is universal in the sense 
that it is independent of the potential 
strength or other microscopic details.  The mobility 
takes a perfect value in the free phase ($\mu=1$), 
vanishes in the localized phase ($\mu=0$), 
and takes a finite value between the two 
in the intermediate phase ($0<\mu<1$).  

This paper is organized as follows.  Sec.\ \ref{sec:overview} 
is devoted to a general and qualitative overview of the 
QBM description of resonant tunneling 
and the multichannel Kondo problem.  In Sec.\ \ref{sec:twoLead}, 
we show how the one-dimensional (1D) 
QBM model is derived from the two-lead quantum 
dot model of 
spinless electrons and give a brief discussion 
on the universal scaling.  We generalize this discussion to 
the multi-lead quantum dot model 
and the multichannel Kondo model 
in Sec.\ \ref{sec:multiLead}.  Sec.\ \ref{sec:luttinger} is 
devoted to the double-barrier 
resonant tunneling model of a Luttinger 
liquid.  We discuss the similarities and differences between 
this model and the quantum dot model.  
In Sec.\ \ref{sec:formalism} 
we develop a general formalism of the QBM model and 
present the dual theories in the weak and strong 
backscattering limits.  
We define the mobility and introduce the renormalization 
group flow equation in both limits.  
In Sec.\ \ref{sec:phase}, we 
apply the QBM theory to specific lattice 
potentials.  The zero-temperature phase 
diagram is derived for each lattice.  It is also shown that 
there is a stable intermediate phase for non-symmorphic lattices.  
In Sec.\ \ref{sec:kondo}, we explicitly show 
that the QBM model is equivalent to the 
Toulouse limit of the multichannel 
Kondo model.  We also show that 
the perturbation calculations of physically relevant 
quantities --- the scaling dimension of the 
leading irrelevant operator 
and the fixed point mobility --- are consistent 
with the exact solutions 
derived from conformal field theory results.  
In Sec.\ \ref{sec:spin1/2}, we calculate the 
on-resonance conductance matrix $G_{ab}$ 
of the multi-lead quantum dot model 
in terms of $\mu$.  Especially for the 
two-lead spin-1/2 problem, we calculate the conventional 
source-drain conductance $G^*$ on resonance, as 
well as the scaling form of the 
resonance line shape.  Finally, 
concluding remarks and discussions 
are given in Sec.\ \ref{sec:conc}.  
Details of some lengthy calculations 
are given in appendices.

\section{Qualitative overview}
\label{sec:overview}

In this section, we will use a few simple 
examples to describe qualitatively the 
mapping between the QBM model, the resonant 
tunneling problem, and the multichannel 
Kondo problem.  Rigorous derivations as 
well as more complicated systems will be 
discussed in the following sections.  

\begin{figure}[tb]
\epsfxsize=1.7in
\centerline{ \epsffile{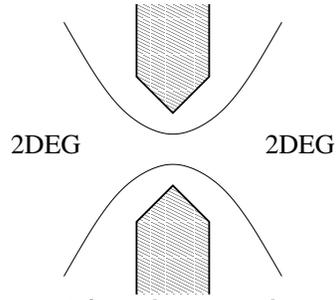} }
\caption{Schematic view of a quantum point contact 
formed in a two-dimensional electron gas (2DEG).  The shaded 
regions are metal gates and the thin lines 
are the boundaries of the 2DEG.  
Negative voltage on the gates depletes electrons 
underneath them and separates the 2DEG into two regions 
connected only by a narrow channel.}
\label{fig:qpc}
\end{figure}

The mapping between the QBM model and 
a mesoscopic tunneling problem 
is most easily understood by considering 
a single QPC formed by gates 
on a two-dimensional electron gas (2DEG), 
which is schematically depicted 
in Fig.\ \ref{fig:qpc}.  
Experiments show that the linear conductance $G$ 
through a QPC is quantized in units of $e^2/h$, 
\cite{wees88,wharam88,timp91inbook} 
which is a characteristic signature of non-interacting 1D 
electron systems. \cite{landauer70}  
Indeed, a QPC may be thought of as 
a wave guide of an electron wave 
function, and each wave mode plays the 
role of a 1D channel.  
Each fully transmitting 1D channel contributes 
one conductance quantum $e^2/h$ to the 
total conductance of a QPC. \cite{matveev95}  

As explained below, it is possible to see 
how a QPC is described by a QBM model, 
without going into too much detail.  
As the first example, we will consider a 
QPC that fully transmits only one 1D channel 
and we will construct a QBM model that reproduces the 
correct quantized conductance.  The current is
obviously given by $I=-e\partial_t Q$, where $Q$ is the number 
of electrons in the source 2DEG lead.  
The Euclidean action may be written as
\begin{equation}
S = S_0 + S_{\text{source}},
\label{eq:SQ}
\end{equation}
where $S_0$ is the unperturbed action, 
and $S_{\text{source}}$ is the source term due to 
voltage drop $V$ across the QPC.  
The second term is straightforwardly 
given from electrostatics by
\begin{equation}
S_{\text{source}} = -\int d\tau eQ(\tau)V,
\label{eq:SSource}
\end{equation}
where $\tau=it$ is the imaginary time.  Using 
the Kubo formula, the linear conductance is written as
\begin{eqnarray}
G & = & \lim_{V\to 0} \frac{\partial}{\partial V} \left< I \right> \\
& = & \lim_{\omega\to 0} \frac{1}{\hbar|\omega|}
  \int d\tau \: e^{i\omega\tau} \left< \text{T}_\tau I(\tau) I(0) \right> \\
& = & \lim_{\omega\to 0} \frac{e^2}{h|\omega|}
  \int d\omega'\:\omega\omega'\left< Q(\omega)Q(-\omega') \right>.
\end{eqnarray}
If we assume that $S_0$ has a simple quadratic form 
and the Hamiltonian does not explicitly depend on time, 
then it {\em must} have the following form in order to 
yield the correct quantized conductance:
\begin{equation}
S_0 = \frac{1}{2} \int d\omega\: |\omega||Q(\omega)|^2.
 \label{eq:S0Q}
\end{equation}
This may be thought of as an action of a QBM 
model written in terms of the spatial 
coordinate of the Brownian particle, $Q$.  

A rigorous derivation of the action 
using the bosonization method \cite{bosonization} 
reveals that the above supposition is 
correct.  In the boson representation, 
a generic non-interacting free 1D electron 
system is described 
by a Euclidean action \cite{haldane81,kane92}
\begin{equation}
S_B = \frac{1}{8\pi v_F} \int d\tau dx \left[ (\partial_\tau \theta)^2 + (v_F \partial_x \theta)^2 \right],
\label{eq:SBoson}
\end{equation}
where $v_F$ is the Fermi velocity.  The boson 
field $\theta(x,\tau)$ may be thought of as the 
local phase of a 1D charge density wave.  
Especially, it satisfies $\partial_t\theta=2\pi I$ 
and $\partial_x\theta=2\pi \rho$, where $I$ is the 
current and $\rho$ is the number density of electrons.  
If we define the coordinate system such 
that $x=0$ at the center of the QPC, then
\begin{eqnarray}
Q(t) & = & \int_0^\infty dx\: \rho(x,t) \\
& = & -\frac{\theta(x=0,t)}{2\pi}+\text{const.}
\end{eqnarray}
Integrating out $\theta(x,\tau)$ except for $x=0$, 
one can easily see that the effective 
action is indeed given by Eq.\ (\ref{eq:S0Q}).  

Note that the non-analytic form of $S_0$ 
suggests that the Brownian particle is not free.  
This is due to the dissipation provided 
by $\theta(x\neq0)$ after they are integrated out.  
The analogy between the QPC problem and the QBM 
model is completed by mapping the current $I$ to the 
average ``final velocity'' of the Brownian particle and 
the voltage drop $V$ to the ``uniform external force''.  
The dissipation is indeed necessary to 
prevent $\partial_tQ$ from increasing indefinitely, 
because we know from the Ohm's law that 
it is finite.  Even though the QBM action may have to 
be modified for more complicated systems, 
the above mappings of fundamental quantities 
are universal.  

\begin{figure}[tb]
\epsfxsize=3.2in
\centerline{ \epsffile{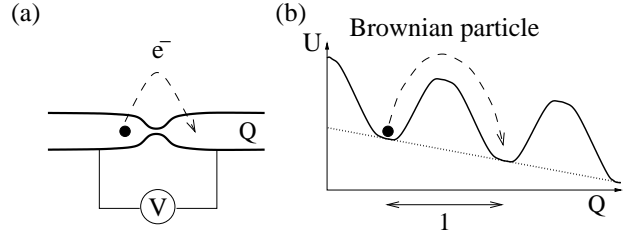} }
\caption{Analogy between a quantum point contact and 
 the quantum Brownian motion in a tilted periodic potential.}
\label{fig:qpcQbm}
\end{figure}

Let us now include backscattering 
by pinching off the point contact via 
gate voltage.  It is equivalent to placing 
a backscattering barrier at the center 
of the QPC that suppresses electron 
tunneling.  If the barrier 
is strong, tunneling is almost suppressed and 
$Q$ is strongly quantized at integers 
due to the discrete nature of electron number.  
On the other hand, if the barrier is weak, 
$Q$ is only weakly quantized.  However, as long as 
the barrier strength is nonzero, quantization 
effect is also nonzero.  This effect is most simply 
taken into account by adding 
a term periodic in $Q$ to the action.  
Since the period is one, the 
most generic backscattering action takes the form
\begin{equation}
S_v = -\sum_{n=1}^\infty \int \frac{d\tau}{\tau_c} \left[ v_n e^{i2\pi nQ(\tau)} + \text{c.c.} \right],
 \label{eq:SvQ}
\end{equation}
where $v_n$ is a dimensionless backscattering 
amplitude and $\tau_c$ is a short-time cutoff.  
In the QBM model, this describes 
a periodic potential.  The potential 
gets tilted if the external uniform force 
term $S_{\text{source}}$ in Eq.\ (\ref{eq:SSource}) 
is also added.  This is schematically illustrated in 
Fig.\ \ref{fig:qpcQbm}.  Transferring 
an electron across the point contact corresponds 
to moving the fictitious Brownian particle from a minimum 
of the potential to the next minimum.  
If we define the ``mobility'' of the 
particle $\mu$ as the final velocity $\left< \partial_tQ \right>$ 
divided by the external force $eV$ in the 
linear response regime, it is obviously 
the same as the electric conductance $G$ 
up to a constant factor $e^2$.  

If the barrier is strong, it is more convenient 
to think of the QPC system as two 
almost separate leads that are weakly coupled 
by a small tunneling amplitude $t$.  
The corresponding QBM model is 
described by hopping of the Brownian particle 
between ``lattice sites'' that lie on deep 
minima of the periodic potential.  In the 
above example, the lattice is a simple 1D lattice 
with lattice constant $a=1$.  

Another example in which the 1D QBM description is 
useful is the interacting 1D 
electron system known as Luttinger 
liquid.  Again, if we focus on the 
current past a certain reference point, say $x=0$, 
we may obtain a QBM model by integrating 
out all degrees of freedom except $Q$, which 
is again defined as 
the total number of electrons on the 
positive axis ($x>0$).  If there is a backscattering 
center such as a point impurity at $x=0$, it gives 
rise to a periodic potential in the QBM model.  
The resulting total action is analogous 
to Eqs.\ (\ref{eq:S0Q}) and (\ref{eq:SvQ}), 
but the Luttinger-liquid correlation modifies 
the coefficient in front of $S_0$ which 
characterizes dissipation.  
Since we can restore the coefficient 1/2 
in Eq.\ (\ref{eq:S0Q}) by rescaling $Q$, 
it may be alternatively said that the Luttinger 
liquid correlation modifies the period of 
the potential.  From this point on, we will 
adopt the convention in which the dissipation 
coefficient is fixed at 1/2 whereas the 
period $a$ is a variable.  

The 1D QBM model was studied 
over a decade ago in quite a 
different context, namely, as a possible description 
of a heavy charged particle 
moving in a metal. \cite{schmid83,fisher85}  Particularly, 
Fisher and Zwerger \cite{fisher85} have used a model of Ohmic 
dissipation proposed earlier by Caldeira and 
Leggett \cite{caldeira83} to show that there are two 
phases at $T=0$: 
(i) For weak dissipation, 
the particle diffuses freely as if the periodic 
potential were completely absent, and the mobility $\mu$ 
takes its maximum perfect value (free phase).  
(ii) When the dissipation exceeds a critical value, 
however, the particle gets localized in a minimum of the 
potential, and $\mu=0$ (localized phases).  
As we adopt the convention in which 
the dissipation coefficient is fixed at 1/2 
[Eq.\ (\ref{eq:S0Q})], the period is given by $a=1$ at 
the phase boundary.  Later, the above result 
was rediscovered in the context of 
single-barrier tunneling in a Luttinger liquid. 
\cite{kane92,furusaki93}  In the Luttinger 
liquid model, the above two zero-temperature phases 
correspond to the two 
different regimes of the electron-electron 
interaction which is typically characterized 
by the Luttinger liquid correlation parameter $g$.  
(i) If the interaction is attractive ($g>1$), 
electrons can freely tunnel through any arbitrarily 
strong barrier (free phase).  (ii) For a repulsive 
interaction ($g<1$), 
however, even an arbitrarily weak barrier 
can completely block transport and the conductance 
vanishes (localized phase).  

Another interesting example that is 
mapped to a 1D QBM model is 
resonant tunneling of spinless 
electrons through a quantum dot.  
In particular, Furusaki and Matveev \cite{furusaki95} have 
recently studied a system of a quantum dot 
that is connected to two leads via QPCs.  
In their model, when an electron tunnels into 
the dot, each electron in the dot has to pay 
the Coulomb energy $\sim e^2/C$, where $C$ 
is the capacitance of the dot.  
If $e^2/C\gg T$, Coulomb blockade 
occurs and the tunneling is suppressed 
unless a nearby gate voltage is tuned 
in such a way that the Coulomb energy 
is exactly canceled by the chemical potential 
of the dot, in which case, resonance occurs.  Another 
assumption of the model is that 
the dot is large enough that the distance between 
the point contacts is much greater than the thermal coherence 
length $L_T=\hbar v_F/k_BT$.  In this limit, an electron 
never travels coherently between the two QPCs.  
Therefore, if we define $Q_1$ and $Q_2$ 
as the number of electrons in the first and the 
second lead, respectively, they 
are independent except through the 
charge conservation condition
\begin{equation}
n + Q_1 + Q_2 = \text{const.},
\end{equation}
where $n$ is the number of electrons in the dot.  
If $e^2/C\gg v_n/\tau_c,T$, 
charge fluctuation in the dot is suppressed and 
we may safely integrate out $n$ (or 
equivalently, $-Q_1-Q_2$).  Then, the resulting 
effective action has the form of a 1D 
QBM action in terms of $Q\equiv Q_2-Q_1$.  The current 
from one lead to the other is $-e\partial_tQ$.  

Resonance is achieved when two charge states 
$n$ and $n+1$ ($n=\text{integer}$) 
have the same energy.  In that case, 
electrons no longer pay the penalty 
when tunneling in and out of the dot.  Thus, 
Coulomb blockade is lifted and 
current is allowed to pass through 
the dot down to $T=0$.  If $T>0$, the conductance is 
finite even off resonance due to thermal 
fluctuations.  However, it decreases as 
the gate voltage is tuned away from resonance.  In Ref.\ 
\ref{ref:furusaki95}, Furusaki and Matveev have exactly 
calculated the resonance line shape of the conductance 
for spinless electrons as a function 
of gate voltage and temperature.  
One of the most interesting features of their result 
is that the resonance line follows a universal 
one-parameter scaling function
\begin{equation}
G(\delta V_G,T) = \tilde{G}\left( \frac{\delta V_G}{T^{1/2}} \right),
\label{eq:GScaling}
\end{equation}
where $\delta V_G$ is the deviation of the gate voltage
from the on-resonance value.  Similar results 
were obtained for resonant tunneling problems 
in a Luttinger liquid \cite{kane92a} or between 
fractional quantum Hall edge states. \cite{moon93,fendley95}  

Furusaki and Matveev have also 
argued that the on-resonance 
behavior of the spinless quantum dot model is governed 
by the fixed point of the two-channel 
Kondo problem on the Emery-Kivelson line. \cite{emery92}  
This limit may be thought of as a generalization of 
the Toulouse limit in the single-channel Kondo 
problem \cite{toulouse70} and may be solved 
exactly. \cite{emery92}  In 
fact, the on-resonance spinless quantum dot 
model may be shown to be {\em equivalent} to a Kondo problem 
on the Emery-Kivelson line.  Specifically, 
it consists of two 1D electron channels 
and a local magnetic impurity, both of which 
have spin 1/2.  
Later, we will rigorously prove the equivalence, 
but it may be rather straightforwardly visualized 
via the following one-to-one mappings: 
(i) the two leads are mapped to the two Kondo channels; 
(ii) the number of electrons $Q_1$ 
and $Q_2$ are mapped to the $z$-components 
of the total spin $S^z_1$ and $S^z_2$ 
in respective Kondo channels; 
(iii) the two degenerate charge states of the 
dot is mapped to the two spin states of the impurity; 
(iv) the tunneling amplitude $t$ is mapped 
to the exchange interaction coupling constant $J$; 
(v) the resonance driving gate voltage 
deviation $\delta V_G$ is mapped to the local magnetic 
field $\mathcal H$ that applies only 
to the impurity spin; 
(vi) the voltage drop between the leads $eV$ 
is analogous to the difference in the external 
magnetic field $\mathcal H_2-H_1$ between the channels; 
(vii) electric conductance $G$ is mapped to the 
``spin conductance'' defined 
as $\lim_{H_2 \to H_1} \partial_t (S^z_2-S^z_1) / ({\mathcal H_2-H_1})$.  

The above Furusaki-Matveev model is quite 
similar to the spinless Kane-Fisher model of 
resonant tunneling through a double-barrier 
in a Luttinger liquid.  \cite{kane92,kane92a}  
However, there are several important differences.  
First of all, the Furusaki-Matveev model assumes no 
interaction in the lead.  Consequently, the 
quasiparticles have charge $-e$ and 
the lattice period of the corresponding QBM model 
is always one.  In the Kane-Fisher model, 
however, Luttinger-liquid 
correlation $g$ can take any value, 
and the potential period changes 
according to $g$.  Another important difference 
is that in the Furusaki-Matveev model, 
the single particle energy level 
spacing $\Delta E$ of the dot is much less than $T$.  
Therefore, the resonance states in the quantum dot 
form a continuum.  
On the contrary, $\Delta E\gg T$ in the Kane-Fisher model 
and the double-barrier structure has only one 
resonance state.  
Consequently, the on-resonance conductance in the 
Kane-Fisher model is given by the maximum perfect 
value for a single channel, $G^*=e^2/h$, whereas 
in the Furusaki-Matveev model, the two QPCs 
behave like two independent resistors in series, 
yielding $G^*=e^2/2h$.  The difference between 
the two models becomes even clearer 
when electrons carry spin.  In Kane-Fisher model, 
the spin excitation in the dot has a gap, which 
is the level spacing $\Delta E$, but 
the spin is gapless in the Furusaki-Matveev model 
since $\Delta E\ll T$.  

So far, we have considered only 1D QBM models, 
where there is only one dynamic variable, namely, $Q$.  
If one needs more than one variable for analysis, 
the above arguments may be extended to higher 
dimensional QBM models.  
Such is the case when the spin or transverse degrees 
of freedom are included into the above two-lead 
quantum dot model, or when the 
quantum dot is coupled to multiple leads.  
As far as low-energy physics is concerned, 
each 1D electron channel 
may be considered equivalent to a lead 
connected to the dot via a QPC with one 
transmitting 1D channel.  
For example, if there are two spin-channels 
in the two-lead model, it is equivalent to 
a four-lead model without spin.  

We may define the number of 
electrons $Q_i\ (i=1\cdots N)$ for each 
individual 1D channel, where 
$N$ is the number of channels.  
The set of all variables, ($Q_1, Q_2, \cdots Q_N$), can be 
thought of as the position 
vector of an $N$-dimensional 
quantum Brownian particle.  In the weak 
coupling limit ($v_n/\tau_c\gg T,e^2/C$), all $Q$'s 
are strongly quantized 
at integers.  The Brownian particle thus 
stays most of the time on $N$-dimensional 
cubic lattice sites with unit 
lattice constant ($a=1$).  
As in the above analysis, the 
number of electrons in the dot, $n=\text{const.}-\sum_iQ_i$, 
may be integrated out at low temperatures ($T\ll e^2/C$).  
In this process, the dimension of the resulting QBM model 
is reduced by one and the cubic lattice is 
projected to the $N-1$ dimensions.  

As the two-lead spinless problem of resonant tunneling 
is mapped to the two-channel Kondo model, 
so does the $N$-channel resonant tunneling problem 
to the $N$-channel Kondo model.  The mapping 
will be performed later in two 
steps.  Firstly, in Sec.\ \ref{sec:multiLead}, we will 
map the $N$-lead resonant 
tunneling problem to a QBM model.  Then, in 
Sec.\ \ref{sec:kondo}, the $N$-channel Kondo 
model will be mapped to the same QBM model.  
As in the two-lead case, the resonance line shape 
follows a universal scaling function, of which 
form may be in principle obtained using the properties of 
the $N$-channel Kondo fixed point.

\section{Resonant tunneling of spinless electrons}
\label{sec:twoLead}

In this section, we will formally construct the 
QBM action for the two-lead spinless 
quantum dot model discussed in the previous section.  
Experimentally, one may completely polarize the spin 
by means of a strong magnetic field.  In order 
to avoid quantum Hall effect, we assume 
that the magnetic field is parallel to the 2DEG plane.  
As in Ref.\ \ref{ref:furusaki95}, 
we will assume $\Delta E \ll T \ll e^2/C$, so quantum 
coherence is thermally destroyed as electrons 
travel across the dot.  This allows us to treat 
the two QPCs independently.  We 
also suppose that the QPCs transmit only one 
1D channel.  Following the standard 
bosonization method, \cite{bosonization} 
we write the Euclidean action as \cite{furusaki95}
\begin{equation}
S = S_0 + S_v + S_C, \label{eq:Scorrug1D}
\end{equation}
where the three terms describe the kinetic energy, 
backscattering, and the Coulomb energy, respectively.  
In our units, $\hbar = k_B = 1$.  

The kinetic term is explicitly given by
\begin{equation}
S_0 = \sum_a \frac{1}{8\pi v_F} \int d\tau dx \left[
 (\partial_\tau \theta_a)^2 + (v_F\partial_x \theta_a)^2
 \right],
\end{equation}
where $\theta_a$ is the boson field of 
the $a$-th lead ($a=1,2$) and $v_F$ is the Fermi velocity of the 
electrons.  In accordance with Ref.\ \ref{ref:furusaki95}, 
each QPC has its own coordinate system.  In particular, 
we set $x=0$ at the center of each QPC and 
assume $x<0$ in the dot and $x>0$ in the leads.  
Backscattering is assumed to occur 
right at $x=0$.  Since the density 
of electrons is given in terms of boson fields by 
\cite{kane92,kane92a,yi96}
\begin{equation}
\rho_a(x) = \frac{\partial_x\theta_a(x)}{2\pi},
\label{eq:rhotheta}
\end{equation}
the total number of electrons in the $a$-th lead 
is given by
\begin{eqnarray}
Q_a & = & \int_0^\infty dx\:\rho_a(x) \\
& = & -\theta_a(0)/2\pi
\end{eqnarray}
up to a constant.  

The backscattering term is given by
\begin{equation}
S_v =  - v \sum_{a=1}^2 \int \frac{d\tau}{\tau_c}
 \left[ e^{i\theta_a(x=0)} + \text{c.c.} \right],
 \label{eq:Sv2lead}
\end{equation}
where $v$ is a dimensionless backscattering amplitude.  
This describes an electron backscattered from 
momentum $k_F$ to $-k_F$ and vice versa 
($2k_F$-backscattering).  In real systems, higher 
harmonics that describe multiple electron 
scattering ($4k_F$-backscattering etc.) are 
also present [See Eq.\ (\ref{eq:SvQ})].  
As will be shown in 
Sec.\ \ref{sec:formalism}, however, 
they are less relevant in the renormalization 
group sense, and it suffices to keep only 
the single-electron backscattering term in order to 
discuss low-energy physics.  We have also assumed that 
the two QPCs are identical and thus have the same backscattering 
amplitude.  The effect of asymmetric QPCs will be discussed 
later at the end of this section.  In general, $v$ is a 
complex number, but without loss of generality it can 
be always made real by shifting $\theta_a$ by a constant.  

Finally, the Coulomb energy term can 
also be written in terms of $\theta_a$.  Since the 
number of electrons in the dot is $n=-(Q_1+Q_2)$ up 
to a constant, we may write
\begin{equation}
S_C = \frac{e^2}{2C} \int d\tau \left[ 
 \frac{\theta_1(x=0)+\theta_2(x=0)}{2\pi} - n_0\right]^2,
\end{equation}
where $n_0$ is proportional to the chemical potential 
of the dot and is controlled by the gate voltage.  
Note that $n_0$ is not integer in general.  

\begin{figure}[tb]
\epsfxsize=1.8in
\centerline{ \epsffile{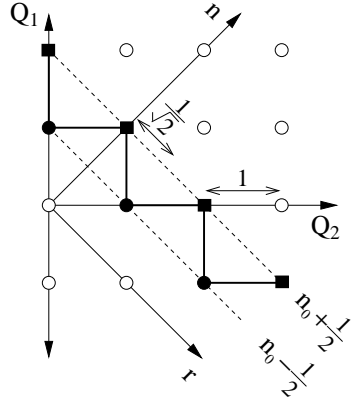} }
\caption{Minima of the periodic potential in $Q_1$-$Q_2$ space.  
 After $n$ has been integrated out, the on-resonance model is 
 described by the corrugated 1D lattice.  The two sets of 
 degenerate sites in the lattice are depicted by 
 filled circles and squares.}
\label{fig:resonance}
\end{figure}

Now, we may derive the QBM action 
by integrating out all degrees 
of freedom but $Q_1$ and $Q_2$.  The resulting 
effective action is written as
\begin{eqnarray}
S_{\text{eff}}[Q_1,Q_2] & = & \frac{1}{2} \int d\omega\:
 |\omega| \left[ |Q_1(\omega)|^2 + |Q_2(\omega)|^2 \right] \nonumber \\
& & -\ 2v \int \frac{d\tau}{\tau_c} [
 \cos 2\pi Q_1(\tau) + \cos 2\pi Q_2(\tau) ] \nonumber \\
& & +\ \frac{e^2}{2C} \int d\tau\: \{-[Q_1(\tau)+Q_2(\tau)]-n_0\}^2.
 \label{eq:Seff2}
\end{eqnarray}
We have replaced the Matsubara frequency sum 
to an integral, which is a valid approximation 
at low temperatures.  
Identifying ($Q_1, Q_2$) with the position vector of 
a Brownian particle, the above action describes 
a two-dimensional (2D) QBM model.  
The second term above is now mapped to a 
periodic potential.  Clearly, it has 
a square lattice symmetry with the 
period $a=1$.  The minima of the periodic 
potential are explicitly depicted by circles 
and squares in $Q_1$-$Q_2$ 
space in Fig.\ \ref{fig:resonance}.  
For weak backscattering, 
one may treat it perturbatively in small $v$.  
In the opposite limit (large $v$), the problem is 
better described in terms of tunneling 
between potential minima with a small 
tunneling amplitude $t$.  This duality 
will be discussed in more detail in Sec.\ \ref{sec:formalism}.  

If $e^2/C\gg T,v/\tau_c$, then $n=-(Q_1+Q_2)$ can be 
safely integrated out.  In terms of the new 
variable $r=(Q_2-Q_1)/\sqrt{2}$, the final 
effective action is written as 
\begin{eqnarray}
S_{\text{eff}}[r] & = & \frac{1}{2}\int d\omega\:|\omega||r(\omega)|^2 \nonumber \\
& & \quad -\ 4v\cos\pi n_0\int \frac{d\tau}{\tau_c} \cos\sqrt{2}\pi r(\tau).
 \label{eq:Seffr}
\end{eqnarray}
Note that the above action has the same form as that of the 
1D QBM model in Sec.\ \ref{sec:overview}, except 
for an additional $n_0$ dependent factor in front 
of the periodic potential and 
a longer period $a=\sqrt{2}$.  
According to Fisher and Zwerger, \cite{fisher85} if $a>1$, 
the mobility $\mu$ vanishes at $T=0$ for any 
arbitrarily small amplitude of the periodic potential.  
In other words, the Brownian particle gets completely 
trapped in a potential 
minimum.  In the original quantum dot problem, this means 
that electron tunneling through the dot is 
completely suppressed at $T=0$, unless $\cos\pi n_0=0$.  
This is the familiar Coulomb blockade.  

If the gate voltage is tuned so that $n_0$ is exactly 
a half integer, the periodic potential is 
identically zero.  This is the resonance condition 
where two charge states $n=n_0\pm 1/2$ become degenerate.  
In the QBM description, this corresponds to a 
degeneracy between two different sets 
of lattice sites in the $Q_1$-$Q_2$ space.  The degenerate 
lattice sites in $Q_1$-$Q_2$ space forms 
a ``corrugated 1D lattice'' as in 
Fig.\ \ref{fig:resonance}.  
Note that the period in the $r$-direction has 
now become $a=1/\sqrt{2}$.  The motion in 
the $n$-direction gets renormalized down 
to zero at $T=0$.  
Since $a<1$, the Brownian particle now diffuses 
freely completely unaffected by the periodic 
potential at $T=0$.  This implies 
that Coulomb blockade is lifted and electron current is 
allowed to flow across the dot.  

We want to make a few comments here.  Firstly, 
the resonance fixed point of the two-lead spinless 
model is the same as the trivial 
backscattering-free fixed point ($v=0$).  
In general, this is not true in higher 
dimensional QBM models, in which the resonance 
fixed point is usually inaccessible using 
a perturbative method.  In this case, however, 
the on-resonance conductance is simply 
given by the perfect conductance calculated 
in the absence of backscattering.  
From the series conductance of two perfect 
conductors, we get $G=e^2/2h$.  

Secondly, the resonance line shape follows a 
scaling function.  
If we define $\delta n_0$ as the deviation of $n_0$ from 
the nearest half integer, one may Taylor expand 
Eq.\ (\ref{eq:Seffr}) for small $\delta n_0$ to get
\begin{equation}
S_v \approx \pm 4\pi v\:\delta n_0 \int \frac{d\tau}{\tau_c} \cos\sqrt{2}\pi r + {\mathcal O}(\delta n_0^3).
\end{equation}
As will be shown in Sec.\ \ref{sec:formalism}, 
the scaling dimension of $\delta n_0$ is 
given by $1/a^2$.  Substituting $a=\sqrt{2}$, 
we find that the conductance 
as a function of $\delta n_0$ and temperature 
has the scaling form 
\begin{equation}
G(\delta n_0, T) = \tilde{G} \left(\frac{\delta n_0}{T^{1/2}}\right).
 \label{eq:GvT}
\end{equation}
This is consistent with the result in 
Ref.\ \ref{ref:furusaki95} 
[See Eq.\ (\ref{eq:GScaling})].  
Clearly, $\tilde{G}(0)=e^2/2h$ and $\tilde{G}(\infty)=0$.
As will be shown in Sec.\ \ref{sec:kondo}, 
if we map this model to a two-channel 
Kondo problem, $\delta n_0$ is mapped to the 
magnetic field $\mathcal H$ that is locally 
applied to the magnetic impurity.  Since the 
scaling dimension of $\mathcal H$ is 
1/2 \nocite{ludwig91} [Ref.\ \ref{ref:ludwig91}], 
it is consistent with the above result.  


Lastly, we briefly discuss the effect of asymmetric QPCs.  
If $v_1\neq v_2$, there is no longer a symmetry 
between $n=n_0\pm 1/2$ even if $n_0$ is a half 
integer.  In fact, the periodic potential never 
vanishes at any $n_0$.  Consequently, even the 
peak value of the conductance resonance vanishes 
as $T\to0$.  In this sense, there is no true resonance 
for asymmetric QPCs.  In the renormalization 
group point of view, the resonance fixed point is 
unstable with respect to the asymmetry.  However, 
even if the QPCs are not perfectly symmetric, 
if $|v_1-v_2|/\tau_c\ll T$, the flow will not have crossed 
over to the asymmetric fixed point and the system 
is still governed by the resonance fixed point.

\section{Multi-lead resonant tunneling models}
\label{sec:multiLead}

In this section, we will generalize the above 
analysis to a general $N$-lead case and show 
that it is mapped to a ($N-1$)-dimensional 
QBM model.  

\begin{figure}
\epsfxsize=3in
\centerline{ \epsffile{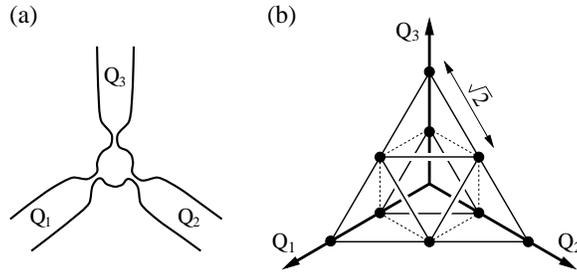} }
\caption{(a) Schematic view of a three-lead quantum dot model.  
(b) Two adjacent planes of triangular lattices, 
or equivalently, a corrugated honeycomb lattice, which are 
embedded in a cubic lattice.  It is a view from the (111) direction.}
\label{threeLead}
\end{figure}

The first example is the three-lead model shown in 
Fig.\ \ref{threeLead}(a).  As in 
the two-lead case, we define the total number of 
electrons in each lead $Q_a$ $(a=1,2,3)$ and 
treat them as vector components of the 
three-dimensional (3D) position 
of a Brownian particle.  Charge conservation law 
again asserts that $n=-(Q_1+Q_2+Q_3)$ up to a constant.  
Assuming the three QPCs are symmetric, 
the total QBM action looks similar to the one 
in Eq.\ (\ref{eq:Seff2}) except for the 
third variable $Q_3$.  Again, in the 
strong backscattering limit ($v/\tau_c\gg T,e^2/C$), $Q_a$ 
are strongly quantized in order to minimize 
the potential energy.  Since the potential has 
a cubic lattice symmetry with 
lattice constant $a=1$, the minima also 
form a cubic lattice [Fig.\ \ref{threeLead}(b)].  
The Brownian particle propagates by 
hopping between these lattice sites.  
At low temperatures ($T\ll e^2/C$), $n$ is 
integrated out and takes an integer 
value that minimizes the Coulomb energy.  
In terms of $Q_a$, this condition is 
\begin{equation}
Q_1+Q_2+Q_3 = -n = \text{integer const.}, 
\end{equation}
which defines a plane perpendicular 
to the (111) direction.  Fig.\ \ref{threeLead}(b) 
shows two examples of such planes 
viewed from the (111) direction.  
On any plane, the subset of lattice sites clearly form 
a triangular lattice with lattice 
constant $a=\sqrt{2}$.  Off resonance, 
$n$ is allowed to take only one value, 
and the Brownian particle stays 
on only one plane.  However, when $n_0$ 
is a half integer (on resonance), 
the system is allowed to hop between two planes 
corresponding to $n=n_0\pm 1/2$.  As sketched in Fig.\ 
\ref{threeLead}(b), they 
form a ``corrugated honeycomb lattice'' 
composed of two parallel triangular sublattice planes.  

We may generalize the above analysis to 
an $N$-lead system.  In the strong 
backscattering limit, the coordinates 
($Q_1, Q_2, \cdots, Q_N$) form an $N$-dimensional 
generalization of a cubic lattice with $a=1$.  When $e^2/C\gg T$, 
depending on whether off or on resonance, 
the Brownian particle is constrained within 
one or two ($N-1$)-dimensional lattices on 
``hyper-planes'' perpendicular to the $(11\cdots1)$ direction.  
It is straightforward to show that the lattice on 
each hyper-plane is an ($N-1$)-dimensional {\em symmorphic} 
\cite{symmorphic} close-packed lattice with $a=\sqrt{2}$.  
We will call it an ``($N-1$)-dimensional 
triangular lattice''.  On resonance 
($n_0=\text{half integer}$), two adjacent hyper-planes 
are degenerate and the two 
($N-1$)-dimensional triangular sublattices are combined 
to form an ``($N-1$)-dimensional honeycomb 
lattice''.  For example, if $N=4$, a 3D triangular 
lattice (off resonance) is, in fact, a face-centered cubic (FCC) 
lattice, \cite{hcp} and a 3D honeycomb lattice (on 
resonance) is a diamond lattice.  Note that a diamond lattice 
is made of two FCC sublattices.  

Now we present a formal derivation of the 
QBM action from the $N$-lead resonant tunneling 
model.  Generalizing Eq.\ (\ref{eq:Seff2}), we 
may write
\begin{eqnarray}
S_{\text{eff}}[\{Q_a\}] & = & \sum_{a=1}^N \left\{
 \frac{1}{2} \int d\omega\: |\omega| |Q_a(\omega)|^2
 - v \int \frac{d\tau}{\tau_c}
 \cos 2\pi Q_a(\tau) \right\} \nonumber \\
& & \qquad +\ \frac{e^2}{2C} \int d\tau
 \left\{ -\left[ \sum_{a=1}^N Q_a(\tau) \right] -n_0 \right\}^2.
\end{eqnarray}
For convenience, we perform change of variables 
\begin{equation}
\biggl[ \begin{array}{c} r_i \\
 \nu \end{array} \biggr]
 = {\mathsf O} \biggl[ \: Q_a \: \biggr], \qquad (i=1\cdots N-1),
 \label{eq:O}
\end{equation}
where $\mathsf O$ is an orthogonal matrix 
which is defined in such a way that 
the last component 
of the left-hand side of the above 
equation is given by
\begin{equation}
\nu \equiv \frac{1}{\sqrt{N}} \sum_a Q_a = - \frac{n}{\sqrt{N}}.
\end{equation}
After $\nu$ is integrated out, the 
effective action becomes
\begin{eqnarray}
S_{\text{eff}}[\{r_i\}] & = &
 \frac{1}{2} \int d\omega\: |\omega| \sum_{i=1}^{N-1}|r_i|^2 \nonumber \\
& & -\ v \sum_{a=1}^N \int \frac{d\tau}{\tau_c}
 \cos 2\pi \left[ \sum_{i=1}^{N-1} {\mathsf O}_{ai}^{-1} r_i(\tau) - \frac{n_0}{N} \right], 
 \label{eq:SeffN}
\end{eqnarray}
which is an ($N-1$)-dimensional 
QBM action.  The minima of the periodic potential 
in $\{r_i\}$ space form an ($N-1$)-dimensional 
triangular lattice unless $n_0$ is a half integer, 
in which case they form an ($N-1$)-dimensional 
honeycomb lattice.  The zero-temperature 
phase diagrams of these lattice models will 
be derived in Sec.\ \ref{sec:phase}.


\section{Resonant tunneling in a Luttinger liquid}
\label{sec:luttinger}

A similar analysis as above may be performed for 
resonant tunneling of a 1D Luttinger liquid 
in a double-barrier structure.  Let us 
first discuss the spinless case.  
Following Ref.\ \ref{ref:kane92a}, we suppose that 
there are $\delta$-function barriers at $x=0$ 
and $x=d$.  The Euclidean action in the 
bosonized form is given by
\begin{equation}
S = \frac{1}{8\pi gv_s} \int d\tau dx \left[
 (\partial_\tau \theta)^2 + (v_s\partial_x \theta)^2 \right]
 + eV_G \int d\tau\: [\theta(d)-\theta(0)]
 +\ v \int \frac{d\tau}{\tau_c}\:
 \{ \cos[\theta(0)] + \cos[\theta(d)] \},
\end{equation}
where $v_s$ is the sound velocity, $g$ is the Luttinger-liquid 
correlation, and $V_G$ is the gate voltage near 
the isolated region $0<x<d$.  If the electron-electron 
interaction is repulsive (attractive), the Luttinger-liquid 
correlation parameter satisfies $g<1$ ($g>1$).  
After $\theta(x)$ is integrated out except at $x=0$ and $d$,  
we get the effective action
\begin{equation}
S_{\text{eff}}[Q,n] = \frac{1}{g} \int d\omega\:|\omega|
 \left[ \frac{|Q(\omega)|^2}{1+e^{-|\omega|d/v_s}}
 + \frac{1}{4} \frac{|n(\omega)|^2}{1-e^{-|\omega|d/v_s}} \right]
 + eV_G \int d\tau\: n(\tau)
 + v\int \frac{d\tau}{\tau_c} \cos 2\pi Q(\tau) \cos\pi n(\tau),
\end{equation}
where
\begin{eqnarray}
Q & = & \frac{\theta(0)+\theta(d)}{4\pi}, \\
n & = & \frac{\theta(d)-\theta(0)}{2\pi}.
\end{eqnarray}
$Q$ is the number of electrons that has transferred 
from one side of the double-barrier to the other 
and $n$ is the number of electrons between the barriers.  

Since there is an energy gap $\Delta E=v_s/gd$ 
for $n(\omega)$ as $\omega\to 0$, we may safely integrate it out 
for $|\omega|\ll\Delta E$.  The low frequency part of the action 
is then written as
\begin{equation}
S_{\text{eff}}[r] =
 \frac{1}{2} \int d\omega\:|\omega||r(\omega)|^2
 + v \cos\pi n_0 \int \frac{d\tau}{\tau_c} \cos 2\pi\sqrt{g}r(\tau),
 \label{eq:SeffKFspinless}
\end{equation}
where $r\equiv Q/\sqrt{g}$ and $n_0=eV_G gd/v_s$.  

The above action closely resembles that 
of the two-lead quantum dot model in Eq.\ (\ref{eq:Seffr}).  
However, there are a few important 
differences.  First of all, 
the potential period is $1/\sqrt{g}$ 
in the Luttinger liquid, 
which varies with the electron-electron 
interaction, whereas it is fixed 
at $\sqrt{2}$ in the quantum dot 
model.  Another important difference is 
that the level spacing $\Delta E$ is finite 
in the Luttinger liquid model.  Thus, 
at low temperatures ($T\ll \Delta E$), 
the thermal coherence length $L_T=\hbar v_s/k_BT$ is much 
greater than the barrier separation $d$.  
In other words, when electrons tunnel through the 
double-barrier, they fully retain their phase 
information.  Therefore, a single boson 
field $\theta(x)$ has been used 
for both barriers, whereas in the quantum dot model, 
each QPC was described by an 
independent field $\theta_a(x)$.  Consequently, the 
two models are not the same even in the non-interacting 
limit of the Luttinger liquid ($g=1$).  In 
that limit, the period is one 
for the Luttinger liquid model and 
the on-resonance conductance is given 
by $G^*=e^2/h$, which differs from that of the 
two-lead quantum dot model by a factor of 2.  
The resonance line shape 
also has a different scaling form,
\begin{equation}
G(\delta n_0,T) = \tilde{G} \left( \frac{\delta n_0}{T^{1-g}} \right).  
\end{equation}
Interestingly, the Luttinger liquid model yields 
the same on-resonance conductance and 
resonance scaling function as the two-lead 
quantum dot model, if $g=1/2$.  

A similar analysis is possible when there 
is spin degree of freedom.  
Now we need two boson fields $\theta_\uparrow$ 
and $\theta_\downarrow$ for up and down 
spin states.  It is convenient to define 
charge and spin fields
\begin{eqnarray}
\theta_\rho & = & \theta_\uparrow + \theta_\downarrow, \\
\theta_\sigma & = & \theta_\uparrow - \theta_\downarrow.
\end{eqnarray}
The effective action is given by
\begin{equation}
S_{\text{eff}}[\{Q_\alpha\},\{n_\alpha\}] = \sum_{\alpha=\rho,\sigma} 
 \frac{1}{g_\alpha} \int d\omega\:|\omega| 
 \left[ \frac{|Q_\alpha(\omega)|^2}{1+e^{-|\omega|d/v_s}}
 + \frac{1}{4} \frac{|n_\alpha(\omega)|^2}{1-e^{-|\omega|d/v_s}}
 \right]
 + \int \frac{d\tau}{\tau_c}\: [ eV_G n_\rho(\tau) + {\mathcal H} n_\sigma(\tau) ] + S_v,
 \label{eq:SeffKFspin1/2}
\end{equation}
where $\mathcal H$ is the magnetic field applied 
{\em only between the barriers} 
and $S_v$ contains all allowed backscattering 
processes.  $g_\rho$ and $g_\sigma$ are 
Luttinger-liquid correlations for the charge and spin 
sectors, respectively.  If there is SU(2) spin 
symmetry, $g_\sigma=2$.  
In the non-interacting limit, $g_\rho = g_\sigma = 2$.  
Physically, $Q_\sigma$ means 
the total spin that has transferred across 
the barriers and $n_\sigma$ means the total spin 
between the barriers (both in units of $\hbar/2$).  
$Q_\rho$ and $n_\rho$ have similar meanings except 
that they are defined for number 
of electrons.  Note that both $n_\rho$ and $n_\sigma$ 
have an energy gap, whereas 
in the quantum dot model, only $n_\rho$ has a gap.  

After both $n_\rho$ and $n_\sigma$ are integrated out, the above 
effective action describes a 2D QBM 
in terms of $Q_\rho$ and $Q_\sigma$.  Particularly 
in the vicinity of $g_\rho=1$ and $g_\sigma=3$, 
Kane and Fisher have found a fixed point similar to the 
intermediate fixed point of the QBM model on a 2D honeycomb 
lattice. \cite{kane92a}  
There, the most 
relevant backscattering terms are
\begin{equation}
S_v = \int \frac{d\tau}{\tau_c}\: [v_e \cos\pi Q_\rho \cos\pi Q_\sigma + v_1 \cos 2\pi Q_\rho + v_2 \sin 2\pi Q_\rho],
\end{equation}
where all $v$'s are real-valued.  
Physically, the first term describes backscattering 
of a single electron and the other two 
backscattering of a spin-singlet pair.  
Defining rescaled variables 
\begin{equation}
r_\rho = Q_\rho / \sqrt{g_\rho}, \qquad r_\sigma = Q_\sigma / \sqrt{g_\sigma},
\end{equation}
the low-frequency effective action is written as
\begin{equation}
S_{\text{eff}}[r_\rho,r_\sigma]
 = \frac{1}{2} \int d\omega \:|\omega| \left[ |r_\rho(\omega)|^2
 + |r_\sigma(\omega)|^2 \right]
 + \int \frac{d\tau}{\tau_c} \left(
 v_e \cos\pi\sqrt{g_\rho}r_\rho \cos\pi\sqrt{g_\sigma}r_\sigma
 + v_1 \cos 2\pi\sqrt{g_\rho}r_\rho
 + v_2 \sin 2\pi\sqrt{g_\rho}r_\rho
 \right). \label{eq:SElngHoneycomb}
\end{equation}
If $g_\sigma/g_\rho=3$, $v_e=2v_1$, and $v_2=0$, 
the above action has a honeycomb lattice 
symmetry in the $r_\rho$-$r_\sigma$ plane.  
For $g_\sigma/g_\rho\neq 3$, the lattice is 
stretched or compressed.  
The fixed point of this ``distorted honeycomb 
lattice'' will be discussed in 
Sec.\ \ref{ssec:strhoney}.

\section{General formalism of the QBM model}
\label{sec:formalism}

In this section, we will present a general 
formalism of the QBM model with a 
periodic potential and its renormalization 
group analysis.  Specific lattice models will be 
discussed in the following section.  
In any arbitrary dimensions a generic QBM action 
is written as
\begin{equation}
S = S_0 + S_v, \label{eq:S0v}
\end{equation}
where
\begin{eqnarray}
S_0 & = & \frac{1}{2} \int d\omega \:|\omega|
 e^{|\omega|\tau_c} |{\mathbf r}(\omega)|^2, \label{eq:S0r} \\
S_v & = & - \int \frac{d\tau}{\tau_c} \sum_{\mathbf G}
 v_{\mathbf G} e^{i 2\pi{\mathbf G}\cdot{\mathbf r}(\tau)}.
 \label{eq:Sv}
\end{eqnarray}
The above two terms represent the 
dissipative kinetic energy and the periodic 
potential, respectively.  $\mathbf r(\tau)$ is 
the trajectory of the Brownian particle as 
a function of imaginary time $\tau$ and $v_{\mathbf G}$ is a 
dimensionless amplitude of the
potential at a reciprocal lattice 
vector $\mathbf G$.  In Eq.\ (\ref{eq:S0r}), 
high frequency fluctuations are exponentially 
suppressed by $\tau_c$.  The potential 
minima form a lattice $\{{\mathbf R}\}$, where
\begin{equation}
{\mathbf G}\cdot{\mathbf R} = \text{integer},
\label{eq:Gdef}
\end{equation}
for any $\mathbf G$ and $\mathbf R$.  
The dissipation is proportional to the 
coefficient in the righthand side of 
Eq.\ (\ref{eq:S0r}), but it may be absorbed 
into $\mathbf G$ by rescaling $\mathbf r$ 
and $\mathbf G$ appropriately.  
As dissipation gets stronger, 
$|{\mathbf R|}$ increases and $|{\mathbf G|}$ decreases.  

Now let us discuss the linear response of 
the Brown particle to a uniform external 
force $\mathbf F$.  Firstly, we add the 
following term to the total action:
\begin{equation}
S_{\mathbf F} = -\int d\tau\: {\mathbf F}\cdot{\mathbf r}(\tau).
\end{equation}
Then the mobility $\mu_{ij}$ is defined 
as the ratio of the average velocity 
of the Brownian particle 
to the external force, in the linear 
response regime.  Following standard linear 
response theory, it is written as
\begin{eqnarray}
\mu_{ij} & \equiv & \lim_{{\mathbf F} \to 0}
 \frac{\partial}{\partial F_j} \langle \partial_t r_i \rangle, \\
& = & \lim_{\omega\to 0} \frac{1}{|\omega|} \int d\omega'\:
 \omega\omega'\langle r_i(\omega) r_j(-\omega') \rangle,
 \label{eq:mur} \\
& = & \mu\delta_{ij}.
\end{eqnarray}
In the last line we assumed that the mobility
is isotropic.  Note that it is normalized 
in such a way that $\mu=1$ (``perfect mobility''), 
if $v_{\mathbf G}=0$.  

In the weak potential limit ($v_{\mathbf G}\ll1$), 
it is natural to construct a perturbation theory 
in small $v_{\mathbf G}$.  
In order to study low-temperature 
properties, we perform a 
renormalization group analysis analogous 
to the ones in Ref.\ \ref{ref:kane92a}.  
The general scheme of the renormalization 
group analysis is described
in Appendix A.  Up to the leading order in $v_{\mathbf G}$,
the renormalization group flow equation 
is given by
\begin{equation}
\frac{dv_{\mathbf G}}{d\ell} = (1 - |{\mathbf G}|^2) v_{\mathbf G}. \label{eq:dvdl}
\end{equation}
Clearly, all $v_{\mathbf G}$ are 
irrelevant if the shortest reciprocal lattice 
vector satisfies $|{\mathbf G}_0|^2<1$.  

On the other hand, if $v_{\mathbf G}\gg1$, 
the Brownian particle is more likely to be localized at a 
minimum $\mathbf R$ of the potential.  In that case, 
transport is dominated by tunneling 
between potential minima, and the 
partition function may be expanded in powers 
of the fugacity of tunneling 
events.  In order to describe tunneling, 
it is more convenient to write the 
action in terms of $\mathbf k(\tau)$, 
which is the trajectory of the particle 
in the momentum space.  The new action, which is dual 
to Eq.\ (\ref{eq:S0v}), is given by
\begin{equation}
S = S_0 + S_t, \label{eq:S0t}
\end{equation}
where
\begin{eqnarray}
S_0 & = & \frac{1}{2} \int d\omega \:|\omega| e^{|\omega|\tau_c} |{\mathbf k}(\omega)|^2, \label{eq:S0k} \\
S_t & = & -\int \frac{d\tau}{\tau_c} \sum_{\mathbf R} t_{\mathbf R} e^{i 2\pi {\mathbf R}\cdot {\mathbf k}(\tau)} \label{eq:St}.
\end{eqnarray}
The two terms represent 
the kinetic energy and the tunneling, 
respectively.  $t_{\mathbf R}$ is the 
tunneling amplitude between two lattice sites
connected by a displacement vector $\mathbf R$.  
The mobility is written as
\begin{equation}
\mu_{ij} =  \delta_{ij} - 
 \lim_{\omega\to 0} \frac{1}{|\omega|} \int d\omega'\:
 \omega\omega'\langle k_{i}(\omega) k_j(-\omega') \rangle.
 \label{eq:muk}
\end{equation}
Since the dual action has exactly same form 
as the original action, it is straightforward to obtain 
the renormalization group flow equation
\begin{equation}
\frac{dt_{\mathbf R}}{d\ell} = (1-|{\mathbf R}|^2) t_{\mathbf R}. \label{eq:dtdl}
\end{equation}

From the above flow equations one can 
easily find two trivial fixed points: 
the localized ($t_{\mathbf R}=0$) and the free 
($v_{\mathbf G}=0$) fixed points.  The mobility 
at the fixed points is simply given by $\mu=0$ 
and $\mu=1$, respectively.  Other fixed points 
will be obtained and analyzed in the next section.

\section{Lattice models and phase diagrams}
\label{sec:phase}

In this section, we study the properties of the 
QBM model with various lattice symmetries.  Using 
a renormalization group analysis, 
we will derive the phase diagram of 
each lattice.  

A lattice is either symmorphic or non-symmorphic 
depending on the crystallographic space group
of the coordinate transformations that leave
the system invariant.  
If the space group is a 
semi-direct product of a pure translation group 
and a pure rotation group, it is called symmorphic.  
Otherwise, it is non-symmorphic. \cite{symmorphic}  For
example, all Bravais lattices are symmorphic, 
and the honeycomb, Kagom{\'e}, and diamond lattices 
are non-symmorphic.  One of the major results of 
this paper is the proof that there is a 
stable intermediate fixed point only 
for non-symmorphic lattices.  
The first two subsections below deal with 
symmorphic lattices and the following three 
with non-symmorphic lattices.

\subsection{One-dimensional lattice}
\label{ssec:QBMoneDim}

The simplest example of a symmorphic 
lattice is a 1D lattice.  According to Eq.\ (\ref{eq:Gdef}) 
the lattice sites in the direct and the reciprocal 
lattice are given by
\begin{eqnarray}
R & = & mR_0, \\
G & = & n/R_0,
\end{eqnarray}
where $R_0$ is the lattice constant, 
and $m$ and $n$ are integers.
It follows from Eqs.\ (\ref{eq:dvdl}) and 
(\ref{eq:dtdl}) that
\begin{eqnarray}
\frac{dv_G}{d\ell} & = & (1 - n^2 / R_0^2) v_G, \quad \qquad (n\neq 0), \\
\frac{dt_R}{d\ell} & = & (1 - m^2 R_0^2) t_R, \qquad (m\neq 0).
\end{eqnarray}

From the above equations, it is obvious 
that if $R_0<1$, all $v_G$ are 
irrelevant, whereas $t_R$ is relevant 
at least for $m=\pm 1$.  Now 
suppose that the bare value of $t_R(m=\pm 1)$ 
is perturbatively small.  
If the system is scaled down to lower energies under 
successive renormalization group 
transformations, $t_R$ will keep growing  
until eventually it departs from 
the perturbative regime.  Presumably, the system will 
continuously flow towards the opposite limit 
where $v_G$ is small.  Under further renormalization 
group transformations, $v_G$ will decrease until 
it vanish at $T=0$.  In more formal words, 
the localized fixed point ($t_R=0$) 
is unstable and the free fixed point ($v_G=0$) 
is stable.  At $T=0$, the Brownian 
particle diffuses completely unaffected by 
the periodic potential.  

In contrast, 
if $R_0>1$, all $t_R$ are irrelevant 
whereas $v_G$ is relevant at least for $n=\pm 1$.  
Now, exactly the opposite happens 
under renormalization group transformations.  
In short, the free fixed point ($v_G=0$) 
is unstable and the localized fixed point ($t_R=0$) 
is stable.  The zero-temperature phase is now localized 
and the particle gets completely 
trapped in a minimum of the potential at $T=0$.  

The physical interpretation of these is as follows.  
For $R_0<1$, the lattice sites are close enough 
for a given dissipation strength 
--- or the dissipation is weak enough for a given 
lattice constant --- that the Brownian particle 
tunnels easily at low temperatures.  
If $R_0>1$, however, 
lattice sites are far apart or the dissipation 
is too strong that the particle can hardly 
tunnel.  In fact, the above 
result is a reproduction of the result of a 1D QBM model 
in Ref.\ \ref{ref:fisher85} and of a single-barrier 
spinless Luttinger liquid model in Ref.\ \ref{ref:kane92a}.  
In the latter model, the Luttinger-liquid correlation 
parameter $g$ is related to the 
lattice constant by
\begin{equation}
g = 1/R_0^2.
\end{equation}
The above free and localized phases 
are equivalent to the fully coherently 
transmitting phase of the Luttinger liquid model 
for attractive interactions ($g>1$, $R_0<1$) and the 
insulating phase for repulsive interactions ($g<1$, $R_0>1$), 
respectively.  

The above result may be summarized 
in terms of the zero-temperature 
universal mobility 
\begin{equation}
\mu^* = \left\{ \begin{array}{ll}
  0 & \text{if\ } G_0<1, \\
  1 & \text{if\ } G_0>1. \\
\end{array} \right.
\end{equation}
Note that the zero-temperature phase is completely 
independent of $v$ or $t$.  The phase 
transition is solely controlled by $R_0$.  

In order to determine the 
zero-temperature phase, it is sufficient to
consider only the shortest lattice 
vectors ${\mathbf R}_0$ and ${\mathbf G}_0$,
because they are the most relevant operators
in the respective perturbative regime.  
Throughout this paper, we will ignore all the other less relevant 
lattice vectors.  However, one can easily include 
less relevant operators when it is necessary.

Without many modifications, the above analysis may 
be applied to square lattices, 
cubic lattices, and their 
higher-dimensional generalization.  
In the vicinity of the phase boundary $|{\mathbf G}_0|^2=1$,
each vector component of $\mathbf r$ or $\mathbf k$ 
gets completely decoupled from the others at $T=0$, 
since all the coupling terms such 
as $e^{i2\pi(r_1+r_2)}$ are irrelevant.  
Therefore, the square lattice and its higher dimensional 
generalizations can be treated as a set of 
independent 1D lattices.

\subsection{Triangular lattice and its $D$-dimensional generalization}
\label{ssec:triangular}

We argued in Sec.\ \ref{sec:multiLead} that 
the off-resonance multi-lead quantum 
dot model is mapped to the QBM model 
on a multi-dimensional triangular lattice.  
In this subsection, we will first 
analyze the 2D triangular lattice model, 
and then generalize it to higher dimensions.  

The reciprocal lattice of a triangular 
lattice is also a triangular lattice.  
We will assume that there is a six-fold symmetry 
and the potential amplitudes are positive
for the six shortest reciprocal 
lattice vectors ${\mathbf G}_0$, i.e.,
\begin{equation}
v_{{\mathbf G}_0}=v=\text{const.}>0.
\end{equation}
From the six-fold symmetry, it also follows that
\begin{equation}
t_{{\mathbf R}_0}=t=\text{const.}
\end{equation}
It is straightforward to see that 
\begin{equation}
|{\mathbf G}_0|^2 |{\mathbf R}_0|^2 = \frac{4}{3}.
\end{equation}
A simple renormalization group analysis as 
in the previous subsection shows 
that if $|{\mathbf G}_0|^2>4/3$ and $|{\mathbf R}_0|^2<1$, 
the free fixed point is stable and 
the localized fixed point in unstable.  
If $|{\mathbf G}_0|^2<1$ and $|{\mathbf R}_0|^2>4/3$, 
the localized fixed point is stable 
and the free fixed point is unstable.  
However, if $1<|{\mathbf G}_0|^2,|{\mathbf R}_0|^2<4/3$, 
{\em both} fixed points are stable and there 
is at least one unstable fixed point 
between them.  Physically, this means 
that there are at least two distinct 
phases at $T=0$: free and localized.  
Although it is very unlikely that there 
are more fixed points, we 
currently do not have an evidence to completely rule 
out the possibility.  
Assuming there are only two stable 
fixed points, the zero-temperature 
phase depends on 
whether the bare value of $v$ is above 
or below the unstable intermediate fixed point.  
As $v$ moves from one side of the unstable 
fixed point to the other, a first order phase 
transition occurs at $T=0$ and the mobility jumps 
between zero and one.  

For small $v$, the intermediate 
fixed point may be perturbatively 
accessed near $|{\mathbf G}_0|^2=1$.  More 
specifically, we introduce a small 
parameter $\epsilon>0$ and perform a perturbative 
renormalization group analysis at $|{\mathbf G}_0|^2=1+\epsilon$.  
Since details of the analysis 
will be given in Appendix A, here we will just 
state the result.  The renormalization group flow 
equation is given by
\begin{equation}
\frac{dv}{d\ell} = -\epsilon v + 2 v^2. \label{eq:dvdlTri}
\end{equation}
Solving $dv/d\ell=0$, we get the intermediate 
fixed point
\begin{equation}
v^* = \frac{\epsilon}{2}.
\end{equation}
The scaling dimension of the 
most relevant operator at the 
fixed point is
\begin{equation}
\Delta=1-\epsilon. \label{eq:dimTri}
\end{equation}
Using Eq.\ (\ref{eq:mur}), we may also 
compute the universal fixed point mobility
\begin{equation}
\mu^* = 1 - \frac{3\pi^2}{2} \epsilon^2. \label{eq:muTri}
\end{equation}
A similar perturbative analysis was 
used in Ref.\ \ref{ref:kane92a}.

The analysis for small $t$ near $|{\mathbf R}_0|^2=1$ 
is very similar.  In fact, since 
both the direct and reciprocal lattices are triangular, 
one may directly use the duality.  Namely,
the small $v$ limit is mapped exactly to the small $t$ limit
if the following substitutions are made:
\begin{eqnarray}
{\mathbf r} & \to & {\mathbf k}, \\
{\mathbf G} & \to & {\mathbf R}, \\
v & \to & t. 
\end{eqnarray}
The mobility $\mu$ is then mapped to $1-\mu$.  
At $|{\mathbf R}_0|^2=1+\epsilon$, the results can be directly read off 
from Eq.\ (\ref{eq:dvdlTri}) through Eq.\ (\ref{eq:muTri}).  
Explicitly, they are
\begin{eqnarray}
& & \frac{dt}{d\ell} = -\epsilon t + 2 t^2, \\
& & t^* = \frac{\epsilon}{2}, \\
& & \Delta = 1-\epsilon, \\
& & \mu^* = \frac{3\pi^2}{2} \epsilon^2.
\end{eqnarray}
Note that there is a self-dual 
point at $|{\mathbf G}_0|^2=|{\mathbf R}_0|^2=\sqrt{3}/2$ 
where the dual models become identical.  
The mobility at the self-dual point is 
obviously given by
\begin{equation}
\mu^*_{\text{sd}} = 1-\mu^*_{\text{sd}} = 1/2.
\end{equation}

\begin{figure}
\epsfxsize=3.2in
\centerline{ \epsffile{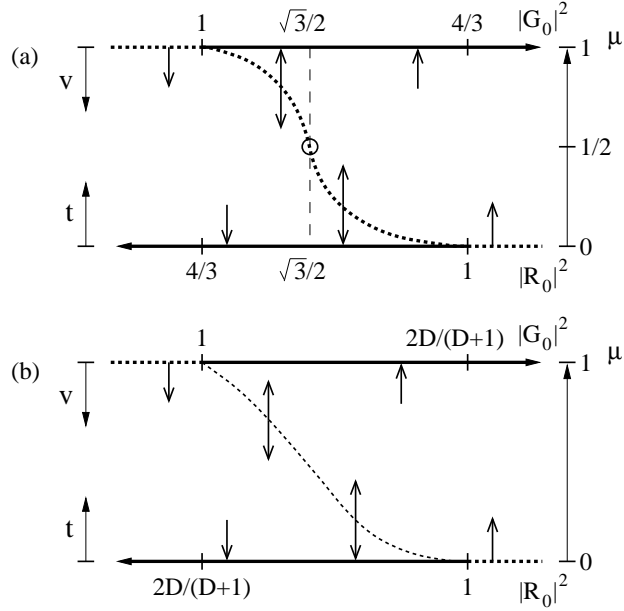} }
\caption{Flow diagrams for symmorphic 
lattices: (a) triangular
lattice; (b) $D$-dimensional generalized 
FCC lattice.  The top (bottom) line in each 
diagram represents the small $v$ ($t$) limit.
Stable (unstable) fixed points
are depicted by solid (dotted) lines, and
arrows indicate the renormalization group flows.  The fixed point
mobility is known exactly at the self-dual 
point (dashed line) and perturbatively 
near $|{\mathbf G}_0|^2=1$ and $|{\mathbf R}_0|^2=1$.}
\label{fig:phaseTrianglular}
\end{figure}

It is very hard to calculate 
the intermediate fixed point 
away from the above two perturbative 
limits and the self-dual point.  
However, assuming that the 
fixed-point lines on both limits smoothly 
join together at the self-dual point, one 
may complete the phase diagram as shown in 
Fig.\ \ref{fig:phaseTrianglular}(a).  
The top and bottom lines of the figure depicts 
the free ($v=0$) and the localized ($t=0$) 
fixed points, respectively.  
The intermediate fixed-point line is 
also drawn along with the renormalization group 
flows.  The above results 
may be summarized in terms of the fixed-point mobility
\begin{equation}
\mu^* = \left\{ \begin{array}{ll}
  0 & \text{if\ } |{\mathbf G}_0|^2 < 1, \\
  0 \text{\ or\ } 1 & \text{if\ } 1 < |{\mathbf G}_0|^2 < 4/3, \\
  1 & \text{if\ } |{\mathbf G}_0|^2 > 4/3. \\
\end{array} \right.
\end{equation}
Unlike in the 1D lattice problem, $v$ and $t$ 
control a first order phase transition, 
if $1 < |{\mathbf G}_0|^2 < 4/3$.  

Now, let us generalize the above 
analysis to the $D$-dimensional 
triangular lattice.  As discussed in 
Sec.\ \ref{sec:multiLead}, a $D$-dimensional 
triangular lattice is defined as a subset 
of the lattice sites of a ($D+1$)-dimensional cubic 
lattice, that are constrained to a hyper-plane 
perpendicular to the $(11\cdots1)$ direction.  
Its lattice sites are equivalent to 
the minima of the off-resonance 
backscattering action of the ($D+1$)-lead 
quantum dot model [Eq.\ (\ref{eq:SeffN}) 
with $n_0\neq\text{half integer}$], except 
for that $|{\mathbf R}_0|$ is no more 
fixed at $\sqrt{2}$. Specifically, it 
is a conventional triangular lattice for $D=2$ and 
an FCC for $D=3$.  
It is hard to visualize an object in 
dimensions greater than three, but it is 
helpful to illustrate 
the $D$-dimensional triangular lattice using 
an induction argument.  For example, 
we can construct an FCC from a 
2D triangular lattice in the following way:  
A triangular lattice is made of periodically 
placed equilateral triangles.  Likewise, an FCC is made of 
equilateral tetrahedrons.  A tetrahedron in an FCC consists of 
a triangle at the bottom and the fourth 
vertex placed above the triangle plane.  Due to periodicity, 
this last vertex also belongs to a triangle in 
another plane parallel to the first.  In other words, 
an FCC is obtained by stacking up planes of triangular lattices.  
However, it has to be done in a certain way that the resulting
lattice is a {\em symmorphic} close-packed structure. \cite{hcp} 
In general, a $D$-dimensional triangular lattice is obtained 
by stacking up ($D-1$)-dimensional triangular 
lattice hyper-planes.  

Now we may perform a similar renormalization 
group analysis as above.  If $D\geq 3$, however, the reciprocal 
lattice \{$\mathbf G$\} is no longer the same as the 
direct lattice \{$\mathbf R$\} and we cannot 
use duality.  (For example, the reciprocal of an FCC 
is a BCC.)  From a geometrical consideration, we find
\begin{equation}
|{\mathbf G}_0|^2 |{\mathbf R}_0|^2 = \frac{2D}{D+1}.
\end{equation}
If $1<|{\mathbf G}_0|^2<2D/(D+1)$, there is 
an unstable fixed point 
that separates the localized and the 
free phases.  The phase diagram is schematically drawn 
in Fig.\ \ref{fig:phaseTrianglular}(b).  
We may also perform an $\epsilon$-expansion analysis 
analogous to the above.  The results are listed 
in Table \ref{tab:v} at $|{\mathbf G}_0|^2=1+\epsilon$ 
in the small $v$ limit.  In the opposite limit of 
small $t$, the results are given in Table \ref{tab:t} 
at $|{\mathbf R}_0|^2=1+\epsilon$.  

Before we move on to non-symmorphic lattices, 
we would like to discuss briefly that 
the symmetry of the lattice 
is crucial on determining the 
phase diagram.  For any arbitrary Bravais 
lattice, $|{\mathbf G}_0| |{\mathbf R}_0| > 1$, 
and obviously, at least one 
of $|{\mathbf G}_0|$ and $|{\mathbf R}_0|$ 
is greater than one.  Therefore, it follows from 
Eqs.\ (\ref{eq:dvdl}) and (\ref{eq:dtdl}) 
that at least one trivial fixed point is stable 
for any arbitrary lattice constant.  
On the contrary, for a non-Bravais 
lattice, $|{\mathbf G}_0|$ and $|{\mathbf R}_0|$ 
may be less than one at the same time, in which 
case both trivial fixed points are 
unstable.  Then, there must 
be a stable intermediate fixed point 
between them.  It is indeed the case 
for the non-symmorphic 
lattices such as the honeycomb and the diamond lattice.  
A common feature of all non-symmorphic lattices is that 
they are made of Bravais sublattices and there 
is a geometrical symmetry between the sublattices.  
Since the asymmetry between sublattices is 
always relevant, even if the symmetry is slightly 
broken, one sublattice will dominate 
the others after repetitive 
renormalization group transformations.  Then, 
the problem reduces to that of the dominating 
sublattice, which is a symmorphic 
lattice.  Therefore, it is the extra symmetry 
between sublattices that prevents this 
``renormalization of basis'', and 
non-symmorphicity is a necessary condition for 
the existence of a stable intermediate fixed point.

\subsection{Honeycomb lattice and its $D$-dimensional generalization}
\label{ssec:honeycomb}

In this and the following two subsections, we will 
study non-symmorphic lattice models.  The first 
example that will be discussed in this subsection 
is the $D$-dimensional honeycomb 
lattice.  Following the framework of the 
the previous subsection, we will first 
consider the conventional 2D honeycomb lattice, 
and then generalize it to higher dimensions.  

\begin{figure}[t]
\epsfxsize=2.8in
\centerline{\epsffile{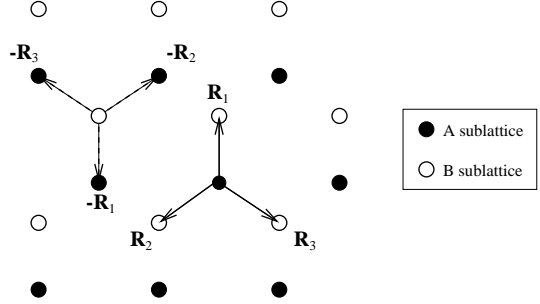}}
\caption{Shortest displacement vectors in a honeycomb lattice.}
\label{fig:honeycomb}
\end{figure}

\begin{figure}[t]
\epsfxsize=3in
\centerline{\epsffile{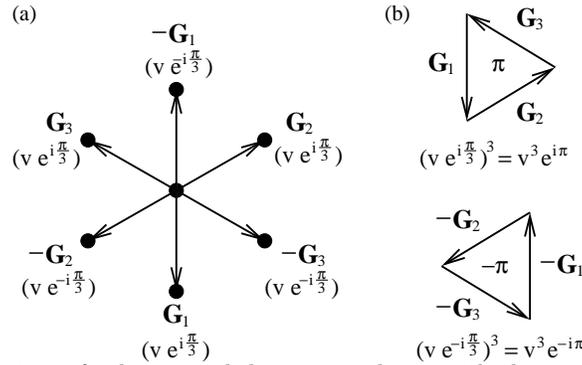}}
\caption{(a) Reciprocal lattice vectors of a honeycomb lattice
are shown with the amplitudes.  (b) A phase factor $e^{\pm i\pi}$ 
is acquired from a trip around a triangle, which is 
analogous to a tight-binding model with flux $\pm\pi$ per 
plaquette.}
\label{fig:tightbinding}
\end{figure}

A honeycomb lattice is made of two 
triangular sublattices.  As depicted in 
Fig.\ \ref{fig:honeycomb}, the 
six shortest lattice vectors ${\mathbf R}_0\in\{\pm{\mathbf R}_a\}$ 
($a=1,2,3$) always connect one sublattice to the other.  
Note that ${\mathbf R}_0\in\{{\mathbf R}_a\}$ on 
A sublattice and ${\mathbf R}_0\in\{-{\mathbf R}_a\}$ on 
B sublattice.  The reciprocal lattice \{$\mathbf G$\} is 
a triangular lattice and the six shortest 
reciprocal lattice vectors are drawn in 
Fig.\ \ref{fig:tightbinding}(a) 
as ${\mathbf G}_0\in\{\pm{\mathbf G}_a\}$ ($a=1,2,3$).  
It is straightforward to see that 
\begin{equation}
|{\mathbf G}_0|^2 |{\mathbf R}_0|^2 = \frac{4}{9}. \label{eq:G2R2Honeycomb}
\end{equation}
From Eqs.\ (\ref{eq:dvdl}) and (\ref{eq:dtdl}), 
it follows that for $4/9<|{\mathbf G}_0|^2<1$, both the free and 
localized fixed points are {\em unstable}.
It implies that there must be at least one 
stable fixed point between the two.  Again, 
we will assume that there is only one stable 
fixed point.  At the new stable fixed point, 
the mobility takes a universal value between 
zero and one ($0<\mu^*<1$).  

Now let us perform an $\epsilon$-expansion analysis 
similar to the one in the previous subsection.  
Computing the structure factor, the 
potential amplitudes are given by
\begin{equation}
v_{{\mathbf G}_0}
 = v(1+e^{i2\pi{\mathbf G}_0 \cdot {\mathbf R}_0})
 = ve^{\pm i\pi/3}, \label{eq:vGHoney1}
\end{equation}
where we have chosen the origin of the 2D space to 
lie on an A-sublattice site.  
The sign in the exponent is ($+$) 
if ${\mathbf G}_0\in\{{\mathbf G}_a\}$ and it is ($-$) 
if ${\mathbf G}_0\in\{-{\mathbf G}_a\}$.  
Using the above equation, the periodic potential 
part of the action may be written as
\begin{eqnarray}
S_v & = & -\int \frac{d\tau}{\tau_c} \sum_{{\mathbf G}_0} v_{{\mathbf G}_0}
 e^{i2\pi {\mathbf G}_0 \cdot {\mathbf r}} \\
& = & -2v \int \frac{d\tau}{\tau_c} \sum_{a=1}^3
 \cos \left( 2\pi {\mathbf G}_a \cdot {\mathbf r} - \frac{\pi}{3} \right).
 \label{eq:SvHoneycomb}
\end{eqnarray}
Since a trip around a triangular plaquette 
in the reciprocal lattice space gives rise to 
a phase $\pm\pi$ 
[Fig.\ \ref{fig:tightbinding}(b)], this problem 
may be also viewed as a tight-binding 
representation of a triangular lattice in which 
a flux $\pi$ is threaded in each plaquette 
with alternating sign.  

The $\epsilon$-expansion analysis in the small $v$ limit
near $|{\mathbf G}_0|^2=1$ is essentially the same as that 
of a triangular lattice except for the extra 
factor $e^{\pm i\pi/3}$ in Eq.\ (\ref{eq:vGHoney1}).  
The results are summarized in Table \ref{tab:v}.  
In the opposite limit of small $t$ near $|{\mathbf R}_0|^2=1$, 
however, the situation is more complicated due 
to the two sublattices.  When many tunneling events occur 
sequentially, ${\mathbf R}_0$ must be alternately chosen 
from $\{{\mathbf R}_a\}$ and $\{-{\mathbf R}_a\}$.  
Technically, this is automatically taken care 
of by using the Pauli matrices $\sigma^\pm = \sigma^x \pm i\sigma^y$.  
The tunneling part of the action may be written as
\begin{equation}
S_t =  -t \int \frac{d\tau}{\tau_c} \sum_a
 \left(\frac{\sigma^+}{2} e^{i2\pi{\mathbf R}_a\cdot{\mathbf k}} + \frac{\sigma^-}{2} e^{-i2\pi{\mathbf R}_a\cdot{\mathbf k}}\right).
 \label{eq:StHoneycomb}
\end{equation}
The alternation between $\{{\mathbf R}_a\}$ 
and $\{-{\mathbf R}_a\}$ is ensured by the identity $(\sigma^\pm)^2=0$.  
Details of the $\epsilon$-expansion calculation 
is given in Appendix A.  The results may be 
summarized as follows:
\begin{eqnarray}
& & \frac{dt}{d\ell} = \epsilon t -  3t^3, \\
& & t^* = \sqrt\frac{\epsilon}{3}, \\
& & \Delta = 1 + 2\epsilon, \\
& & \mu^* = \pi^2 \epsilon.
\end{eqnarray}

\begin{figure}
\epsfxsize=3.2in
\centerline{ \epsffile{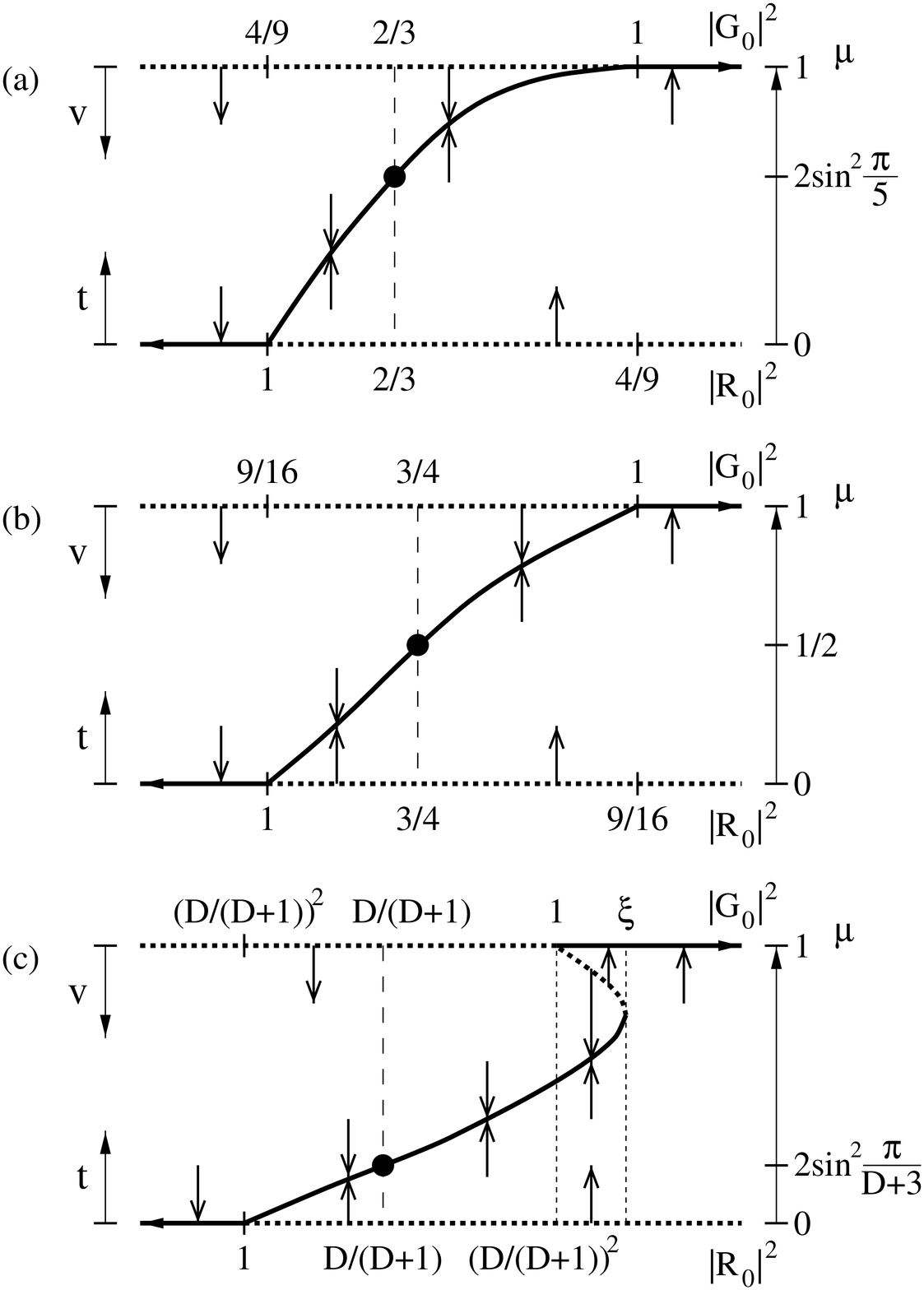} }
\caption{Flow diagrams similar to 
Fig.\ 4 for non-symmorphic
lattices -- (a) honeycomb
lattice, (b) diamond lattice, and (c) their
generalizations to $D\geq 4$.  The fixed
point mobility is known exactly at the 
Toulouse limit (dashed lines) and perturbatively
near $|{\mathbf G}_0|^2=1$ and $|{\mathbf R}_0|^2=1$.}
\label{fig:phaseHoneycomb}
\end{figure}

Fig.\ \ref{fig:phaseHoneycomb}(a) shows the 
phase diagram of the honeycomb lattice.  
Since the stable intermediate fixed-point line 
smoothly connects the localized and the 
free phases, there is no first order phase 
transition as in symmorphic lattices.  
The fixed point mobility $\mu^*$ is 
also a continuous function of $|{\mathbf G}_0|^2$.  
In general, the calculation of $\mu^*$ is not easy 
for an arbitrary $|{\mathbf G}_0|^2$, but as will
be shown in the Sec.\ \ref{sec:kondo}, an exact 
solution may be obtained for the 
special case where $|{\mathbf G}_0|^2 = 2/3$, 
using the mapping of the honeycomb lattice 
model to the Toulouse limit of the three-channel 
Kondo model.  

Now let us generalize the above analysis 
to $D\geq 3$.  First of all, we define a $D$-dimensional 
honeycomb lattice as a subset of the lattice sites 
of a ($D+1$)-dimensional cubic lattice, that 
lie on two adjoining hyper-planes 
perpendicular to the $(11\cdots1)$ direction.  
Its lattice sites coincide with the minima 
of the on-resonance 
backscattering action of the ($D+1$)-lead 
quantum dot model [Eq.\ (\ref{eq:SeffN}) 
with $n_0=\text{half integer}$], except for 
that $|{\mathbf R}_0|$ is no longer fixed.  
For example, a ``3D honeycomb 
lattice'' is a diamond lattice.  
As a honeycomb lattice consists of two triangular 
sublattices and a diamond lattice consists of two 
interpenetrating FCC sublattices, 
a generic $D$-dimensional honeycomb lattice consists 
of two interpenetrating $D$-dimensional triangular 
sublattices.  Among the two basis sites of a $D$-dimensional 
honeycomb lattice, if one is placed at a vertex of 
a $D$-dimensional equilateral $(D+1)$-hedron, the other 
is at the center.  

Generalizing Eqs.\ (\ref{eq:G2R2Honeycomb}) through 
(\ref{eq:StHoneycomb}), we get
\begin{eqnarray}
& & |{\mathbf R}_0|^2 |{\mathbf G}_0|^2 = \left( \frac{D}{D+1}\right)^2, \\
& & v_{{\mathbf G}_0} = ve^{\pm i\pi/(D+1)}, \\
& & S_v = -2v \int \frac{d\tau}{\tau_c} \sum_{a=1}^{D+1}
 \cos \left( 2\pi {\mathbf G}_a\cdot {\mathbf r}
 - \frac{\pi}{D+1} \right), \label{eq:SvDHoneycomb} \\
& & S_t = -t \int \frac{d\tau}{\tau_c} \sum_{a=1}^{D+1}
 \left(\frac{\sigma^+}{2} e^{i2\pi{\mathbf R}_a\cdot{\mathbf k}}
 + H.c. \right). \label{eq:StDHoneycomb}
\end{eqnarray}
In terms of the matrix $\mathsf O$ defined 
in Eq.\ (\ref{eq:O}), the lattice vectors 
are explicitly given by
\begin{eqnarray}
{\mathbf G}_a & = & c({\mathsf O}^{-1}_{a1}, {\mathsf O}^{-1}_{a2}, \cdots, {\mathsf O}^{-1}_{aD}), \label{eq:Ga} \\
{\mathbf R}_a & = & -\frac{1}{c}({\mathsf O}^{-1}_{a1}, {\mathsf O}^{-1}_{a2}, \cdots, {\mathsf O}^{-1}_{aD}) \nonumber \\
& = & -\frac{{\mathbf G}_a}{c^2} \qquad (c=\text{const.}).
\label{eq:Ra}
\end{eqnarray}
Note that Eq.\ (\ref{eq:SvDHoneycomb}) is identical 
to the second term in Eq.\ (\ref{eq:SeffN}) 
if $N=D+1$, $n_0=-1/2$, and $c=1$.  
There is a stable intermediate fixed point 
if $[D/(D+1)]^2<|{\mathbf G}_0|^2<1$.  The results
of the $\epsilon$-expansion analysis are summarized in 
Tables \ref{tab:v} and \ref{tab:t}.  

Quite interestingly, the phase diagrams for $D\leq3$
[Fig.\ \ref{fig:phaseHoneycomb}(a),(b)] and for $D\geq 4$
[Fig.\ \ref{fig:phaseHoneycomb}(c)] have quite different 
structures.  Unlike for $D=2$ or $3$, the 
fixed-point line for $D\geq 4$ approaches the 
top line from right.  Consequently, arbitrarily 
close to the top line, the intermediate fixed points 
are {\em unstable}.  As schematically shown 
in Fig.\ \ref{fig:phaseHoneycomb}(c), 
we conjecture that the fixed-point 
line eventually curves back at a certain 
critical point $\xi$ and gets continuously connected 
to the bottom line.  Unfortunately, 
this critical point is not perturbatively 
accessible, so a definitive evidence to 
prove (or disprove) the above conjecture is currently 
not available.  In the next subsection, however, 
we present an argument that supports 
the conjecture, by using an analysis on 
a distorted diamond lattice.  For this 
exotic model, the direction from which 
the intermediate fixed-point line approaches 
the top line changes {\em continuously} from 
right to left as the lattice is deformed.  
Thus, if the fixed-point line 
is to be smoothly connected to the bottom line 
for $D\leq3$, it is very likely that it will also be the 
case for $D\geq4$ despite the difference, because 
the different feature (the approaching angle) 
is continuously deformable.  
In short, the above conjecture 
is no less reasonable for $D\geq4$ than for $D\leq3$.  
As shown in Fig.\ \ref{fig:phaseHoneycomb}(c), 
if $1<|{\mathbf G}_0|^2<\xi$, the free fixed point 
and one of the two intermediate fixed 
points are both stable.  A first order 
phase transition between the free and the 
intermediate phases 
occurs at $T=0$ as $v$ or $t$ is tuned.

\subsection{Distorted diamond lattice}

In this section, we will deform a diamond lattice 
in a continuous manner to see how the angle 
with which the intermediate fixed line approaches 
the free fixed-point line changes.  
A distorted diamond lattice is defined by 
the minima of the potential
\begin{equation}
V = -2\frac{v}{\tau_c} \sum_{a=1}^4 \cos\left( 2\pi{\mathbf G}_a\cdot{\mathbf r} - \frac{\pi}{4} \right),
\end{equation}
where
\begin{eqnarray}
{\mathbf G}_1 & = & c (\alpha,1,1), \nonumber \\
{\mathbf G}_2 & = & c (\alpha,-1,-1), \nonumber \\
{\mathbf G}_3 & = & c (-\alpha,1,-1), \nonumber \\
{\mathbf G}_4 & = & c (-\alpha,-1,1), \qquad (c, \alpha=\text{const.}).
\end{eqnarray}
This lattice is obtained by 
stretching (or compressing) a diamond lattice in 
the (100) direction.  The regular diamond 
lattice is restored when $\alpha=1$.  

One of the shortest displacement vectors 
connecting the two sublattices is 
\begin{equation}
{\mathbf R}_1 = -\frac{1}{4c} \left( \frac{1}{\alpha},1,1 \right).
\end{equation}
Thus, the relation between the lengths of 
the shortest direct and reciprocal lattice 
vectors may be written
\begin{equation}
|{\mathbf G}_0|^2 |{\mathbf R}_0|^2 = \frac{(\alpha^2+2)(2\alpha^2+1)}{16\alpha^2}.
\end{equation}
If $(11-\sqrt{105})/4 < \alpha^2 < (11+\sqrt{105})/4$, 
then $|{\mathbf G}_0|^2 |{\mathbf R}_0|^2<1$ and 
both trivial fixed points are unstable.  
Using a similar $\epsilon$-expansion calculation as 
in the previous subsection, it is straightforward 
to see that there is a stable intermediate 
fixed point at $|{\mathbf R}_0|^2=1-\epsilon$ near the 
localized fixed point ($\epsilon>0$).  The derivation 
of a flow equation at $|{\mathbf G}_0|^2=1-\epsilon$ 
for small $v$ is more complicated, but it may 
be eventually written as
\begin{equation}
\frac{dv}{d\ell} = \epsilon v - f(\alpha) v^3.
\end{equation}
The function $f(\alpha)$ is computed 
numerically.  If $\alpha<1.9073$, then $f(\alpha)>0$ 
and the solution to the above equation exists only 
if $\epsilon>0$.  The intermediate fixed points 
are stable and approach the top 
line from left, as is the case for a 
regular diamond lattice [Fig.\ \ref{fig:phaseHoneycomb}(b)].  
On the other hand, if $\alpha>1.9073$, then $f(\alpha)<0$ 
and the solution exists only if $\epsilon>0$.  The 
fixed points are now unstable and approach the top 
line from right.  Consequently, the 
phase diagram resembles that of a four or higher 
dimensional honeycomb lattice 
[Fig.\ \ref{fig:phaseHoneycomb}(c)].  
Since $f(\alpha)$ is a continuous function, 
the angle with which the fixed-point line 
approaches the top line is also 
continuous.  Therefore, 
we come to the following conclusion: If our 
conjecture in the previous subsection, that 
the two trivial fixed-point lines are connected 
by a single intermediate 
fixed-point line, is correct for $D\leq3$, 
it is most likely correct for $D\geq4$, too, 
even though the intermediate fixed-point line 
has a different shape in the two cases.

\subsection{Distorted honeycomb lattice: resonant 
tunneling of a spin-1/2 Luttinger liquid}
\label{ssec:strhoney}

As argued in Sec.\ \ref{sec:luttinger}, the 
double-barrier resonant tunneling model of 
spin-1/2 Luttinger liquid 
is mapped to the distorted honeycomb lattice.  
Following Ref.\ \ref{ref:kane92a}, we will 
briefly discuss the intermediate fixed point 
of this exotic lattice problem.  

\begin{figure}
\epsfxsize=1.8in
\centerline{ \epsffile{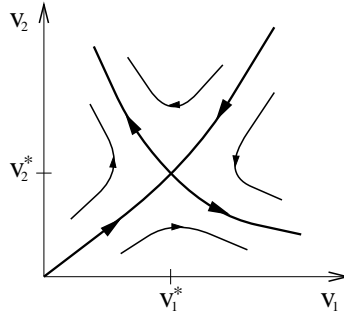} }
\caption{Schematic diagram of renormalization group flows for the
distorted honeycomb lattice model, which describes 
resonant tunneling of spin-1/2 Luttinger liquid [Ref.\ \ref{ref:kane92a}].}
\label{fig:RGDistHoneycomb}
\end{figure}

Since all $v$'s in Eq.\ (\ref{eq:SElngHoneycomb}) 
are marginal if $g_\rho=1$ and $g_\sigma=3$, 
one can perform a perturbative 
renormalization group analysis 
at $g_\rho=1-\epsilon_\rho$ and $g_\sigma=3-\epsilon_\sigma$, for small positive $\epsilon$'s.  
According to Ref.\ \ref{ref:kane92a}, 
there is a stable intermediate fixed point 
at $v_e^*=\sqrt{\epsilon_\rho(\epsilon_\rho+\epsilon_\sigma)}$,  $v_1^*=(\epsilon_\rho+\epsilon_\sigma)/4$, and $v_2^*=0$.  
The renormalization group flow diagram is schematically 
drawn in Fig.\ \ref{fig:RGDistHoneycomb} in the $v_e$-$v_1$ 
plane, assuming $v_2=0$.  
This intermediate fixed point is a continuation of 
that of the regular honeycomb lattice.  However, 
unlike the regular honeycomb lattice, which is 
described by only one parameter $v$, 
there are now three parameters.  Since 
the intermediate fixed point is stable 
only in one direction, in order to obtain 
resonance, one has to tune not only one, 
but two parameters, for example, $v_2$ and $v_e/v_1$.

\section{Multichannel Kondo problem}
\label{sec:kondo}

In this section, we will explicitly derive the 
mapping between the multichannel Kondo model and 
the QBM model.  Then using the mapping, we 
will exactly calculate the mobility $\mu^*$ 
at the intermediate fixed point.  

In the absence of a magnetic field, 
the Hamiltonian of the $N$-channel Kondo model is
given by \cite{emery92,ludwig91} 
\begin{equation}
H = H_0 + H_J,
\end{equation}
where
\begin{eqnarray}
H_0 & = & i v_F \sum_{a,s} \int dx\:\psi_{as}^\dagger\partial_x\psi_{as}, \label{eq:H0Kondo} \\
H_J & = & 2\pi v_F \sum_a \biggl\{ J_z S^z_{\text{imp}} s_a^z(0) \nonumber \\
& & + \frac{1}{2} J_\perp \left[ S^+_{\text{imp}} s_a^-(0) + \text{H.c.} \right]
 \biggr\}. \label{eq:HJKondo}
\end{eqnarray}
The first term $H_0$ is the kinetic energy 
of electrons and $H_J$ is 
the exchange coupling between the impurity 
spin $\vec S_{\text{imp}}$ and the sum of electron 
spins $\sum_a\vec s_a = \sum_a\psi^\dagger_{as} (\vec\sigma_{ss'}/2) \psi_{as'}$ 
at $x=0$.  $\vec\sigma$ is the Pauli matrix and $a$ and $s$ 
are channel and spin indices, respectively.  For generality, 
we assume $J_z$ and $J_\perp$ are independent.  Following 
Emery and Kivelson, \cite{emery92} we bosonize 
the electron operators
\begin{equation}
\psi_{as} = \frac{1}{\sqrt{2\pi v_F \tau_c}} e^{-i\phi_{as}},
\end{equation}
where $v_F$ is the Fermi velocity, $\tau_c$ is the 
short-time cutoff, and $\phi_{as}$ is a boson 
field that satisfies the following 
commutation relation
\begin{equation}
\left[ \phi_{as}(x), \phi_{a's'}(x') \right] = -i\pi\delta_{aa'}\delta_{ss'}\:\text{sgn}(x-x').
\end{equation}
If we separate spin $\phi^\sigma_a=\phi_{a\uparrow}-\phi_{a\downarrow}$ from 
charge $\phi^\rho_a=\phi_{a\uparrow}+\phi_{a\downarrow}$, we may write $\vec s_a$ as
\begin{eqnarray}
s^z_a & = & \frac{1}{4\pi} \partial_x\phi^\sigma_a, \\
s^\pm_a & = & \frac{1}{2\pi v_F\tau_c} e^{\pm i\phi^\sigma_a}.
\end{eqnarray}
In the boson representation, 
the Kondo Hamiltonian is written as
\begin{eqnarray}
H_0 & = & \sum_{a=1}^N \frac{v_F}{8\pi} \int dx \Bigl[
 (\partial_x \phi^\rho_a)^2 + (\partial_x \phi^\sigma_a)^2 \Bigr], \\
H_J & = & \frac{1}{2} \sum_{a=1}^N
 \left\{ v_FJ_zS^z_{\text{imp}} \partial_x\phi^\sigma_a(0)
 + \frac{J_\perp}{\tau_c}\left[ S^+_{\text{imp}}
 e^{-i\phi^\sigma_a(0)}
 + \text{H.c.} \right] \right\}. \label{eq:HJ}
\end{eqnarray}
Since $\phi^\rho_a$ is free and non-interacting, we will 
drop it from now on.  

The first term in Eq.\ (\ref{eq:HJ}) may be eliminated 
by introducing a unitary transformation
\begin{equation}
{\mathsf U}_\lambda
 = e^{i\lambda S^z_{\text{\tiny imp}}\sum_a\phi^\sigma_a(0)},
\end{equation}
which rotates the axes of the spin space at $x=0$.
The Hamiltonian transforms as
\begin{eqnarray}
H' & = & {\sf U} H {\sf U}^{-1} \\
& = & \sum_a \biggl\{ \frac{v_F}{8\pi} \int dx 
 (\partial_x\phi^\sigma_a)^2
 + v_F \left(\frac{J_z}{2} - \lambda\right)
 S^z_{\text{imp}} \partial_x\phi^\sigma_a(0)
 + \frac{J_\perp}{2\tau_c} 
 \left[ S^+_{\text{imp}} e^{-i \left[ \phi^\sigma_a(0) -
 \lambda\sum_b\phi^\sigma_b(0) \right]} + \text{H.c.} \right] \biggr\}.
 \label{eq:H'}
\end{eqnarray}
Now, we set $\lambda=J_z/2$ so that the second term 
above vanishes.  A similar technique has been 
used in Ref.\ \ref{ref:emery92} for the two-channel 
Kondo problem and in \nocite{fabrizio94} Ref.\ \ref{ref:fabrizio94} 
for the four-channel problem.  

For later convenience, we make change of variables using 
the the same orthogonal transformation $\mathsf O$ 
as in Eq.\ (\ref{eq:O}): 
\begin{equation}
\left[ \left. \left. \begin{array}{c} \phi_i^{\text{sf}} \\
 \phi^{\text{s}} \end{array} \right]
 = {\mathsf O} \right[ \: \phi^\sigma_a \: \right], \qquad (i=1\cdots N-1).
\end{equation}
The last component of the left-hand 
side, $\phi^{\text{s}} = \sum_a \phi^\sigma_a/\sqrt{N}$ 
is proportional to the sum of the spins in all channels, 
whereas the other components $\phi_i^{\text{sf}}$ 
($i=1, \cdots, N-1$) are ``spin flavors''.
In terms of these new variables, the above Hamiltonian
finally becomes
\begin{equation}
H' = \frac{v_F}{8\pi} \int dx
 \left[ (\partial_x\phi^{\text{s}})^2 + \sum_{i=1}^{N-1}
 (\partial_x\phi_{i}^{\text{sf}})^2 \right]
 + \frac{J_\perp}{2\tau_c} \sum_{a=1}^N \left\{
 S^+_{\text{imp}} e^{-i \left[ \frac{1}{\sqrt{N}}
 \left(1-\frac{N}{2}J_z\right)\phi^{\text{\tiny s}}(0)
 + \sum_{i=1}^{N-1} {\mathsf O}^{-1}_{ai}\phi_{i}^{\text{\tiny sf}}(0) \right]}
 + \text{H.c.} \right\}. \label{eq:HKondoBoson}
\end{equation}
Note that 
there is a reflection symmetry about $J_z=2/N$.  
More specifically, $H'$ is invariant under 
$\phi^{\text{s}}\to -\phi^{\text{s}}$ and $J_z\to -J_z+4/N$.  
At the symmetry point $J_z=2/N$, the total spin $\phi^s$ 
gets completely decoupled and becomes non-interacting.  
This is the multichannel generalization of 
the Toulouse limit of the single-channel 
model \cite{toulouse70} or the Emery-Kivelson line 
of the two-channel model. \cite{emery92}  Although 
this special limit is solved exactly for $N\leq 2$, 
it is not the case for $N\geq 3$.  
As will be shown later, however, there is a 
stable fixed point in the Toulouse limit, 
so some of the low-energy properties may be 
understood via a renormalization group analysis.  
This fixed point is identical to the stable intermediate 
fixed point of the QBM model on an ($N-1$)-dimensional 
honeycomb lattice with $|{\mathbf G}_0|^2=(N-1)/N$.  

Now we may obtain the QBM action using 
the same technique that we have used for 
the multi-lead quantum dot model.  
By setting $J_z=2/N$ and integrating out 
all degrees of freedom away from $x=0$, we get 
\begin{equation}
S_{\text{Kondo}} = \frac{1}{8\pi^2} \int d\omega\:
 |\omega|\sum_{i=1}^N |\phi^{\text{sf}}_i(\omega)|^2
 + \frac{J_\perp}{2} \int \frac{d\tau}{\tau_c} \sum_{a=1}^N
 \left[ S_{\text{imp}}^+ e^{-i\sum_{i=1}^{N-1}
 {\mathsf O}^{-1}_{ai}\phi^{\text{\tiny sf}}_i(0)} + \text{c.c.} \right].
\end{equation}
The above action has {\em identically} the same form as 
the multi-dimensional honeycomb lattice model in 
Eqs.\ (\ref{eq:S0k}), (\ref{eq:StDHoneycomb}), 
and (\ref{eq:Ra}), with $c=1$.  
The mapping becomes complete if we make 
the following substitutions:
\begin{eqnarray}
N & \leftrightarrow & D+1,  \label{eq:mapKondoQBMstart} \\
\phi^{\text{sf}}_i(x=0) & \leftrightarrow & 2\pi k_i, \\
J_\perp & \leftrightarrow & 2t, \\
S^\pm_{\text{imp}} & \leftrightarrow & \sigma^\pm/2, \\
{\mathsf O}^{-1}_{ai} & \leftrightarrow & {\mathbf R}_0.
 \label{eq:mapKondoQBMend}
\end{eqnarray}

This mapping may also be understood using 
the ``spin configuration lattice'', just as 
we have used lattices in $Q_1-Q_2-\cdots-Q_N$ space 
for the quantum dot problems.  
The spin configuration lattice lives 
in a space where each axis represents 
the total spin in channel $a$, 
\begin{equation}
S^z_a = \int dx\: s^z_a(x).
\end{equation}
For small coupling $J_\perp$, the change 
in $S^z_a$ will occur in a step-like manner 
with unit step height ($\Delta S^z_a=1$).  For the $N$-channel 
model, therefore the vectors 
($S^z_1, S^z_2, \cdots$) form an $N$-dimensional cubic 
lattice with lattice constant $a=1$.  
In this lattice, a spin transfer between 
channels is represented by tunneling of a 
fictitious particle between lattice sites.  
The conservation condition of the total electron spin 
{\em plus} the impurity spin 
gives a constraint in the lattice space.  
Explicitly, the constraint 
may be written as
\begin{equation}
\sum_a S_a^z = \text{const.}-S_{\text{imp}}^z.
\end{equation}
This is an equation that defines 
an ($N-1$)-dimensional hyper-plane perpendicular 
to ($11\cdots1$) direction.  Since $S^z_{\text{imp}}=\pm 1/2$, 
the fictitious particle is allowed to 
reside on two adjoining hyper-planes that 
satisfy the above condition.  For the three-channel 
Kondo model, the two (111) planes are 
analogous to those shown in 
Fig.\ \ref{threeLead}(b).  In the Toulouse 
limit ($J_z=2/N$), the motion perpendicular 
to the planes gets renormalized at low temperatures 
and the two planes eventually collapse onto each other at $T=0$.  
As a result, we end up with a new tunneling problem in 
a single plane.  In the three-channel 
case, the resulting lattice is a 2D honeycomb 
lattice [Fig.\ \ref{threeLead}(b)].  In general, 
the $N$-channel Kondo problem in the Toulouse limit 
is mapped to an $(N-1)$-dimensional honeycomb 
lattice QBM model.  

\begin{figure}
\epsfxsize=2.8in
\centerline{ \epsffile{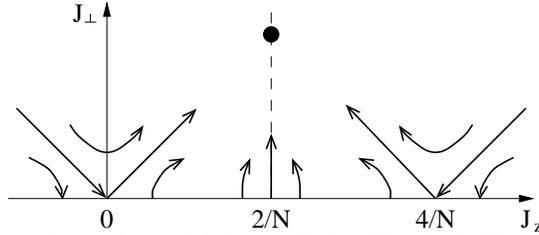} }
\caption{Flow diagram of the $N$ channel Kondo model for
small $J_\perp$.  The dashed line is the Toulouse
limit, $J_z=2/N$.  The strong coupling fixed point is marked
with the full circle.  }
\label{fig:RGKondo}
\end{figure}

A renormalization group analysis 
similar to that of Anderson, 
Yuval, and Hamann \cite{anderson70} 
may be performed to obtain the 
flow diagram of the multichannel Kondo problem.  
Details of the calculations are given 
in Appendix B and we will only present 
the result here:
\begin{eqnarray}
\frac{dJ_z}{d\ell} & = & J_\perp^2 ( 1 - \frac{N}{2} J_z ),
 \label{eq:dJzdl} \\
\frac{dJ_\perp}{d\ell} & = & J_z J_\perp
 ( 1 - \frac{N}{4} J_z ) - \frac{N}{4}J_\perp^3. \label{eq:dJpdl}
\end{eqnarray}
The above equations are exact in $J_z$ and 
perturbative in $J_\perp$.  It is also exact 
in $N$, except for the last term in 
Eq.\ (\ref{eq:dJpdl}), in which we kept only the 
leading order term in $1/N$.  The renormalization 
flow diagram is schematically drawn in 
Fig.\ \ref{fig:RGKondo}.  Obviously, 
the multichannel Kondo fixed point lies 
on the Toulouse line $J_z=2/N$.  Although it is a highly
non-trivial strong coupling fixed point, 
we may use known results of the conformal 
field theory to exactly compute some physical 
quantities.  On the other hand, we may 
use the mapping to the QBM model and 
perturbatively access the fixed point in 
the large $D$ limit of the $D$-dimensional 
honeycomb lattice model.  Therefore, it is 
possible to directly compare the two results, 
in order to test the mapping and the perturbative 
analysis of the QBM model in a non-trivial way.  
Specifically, 
we will calculate below the scaling dimension of 
the leading irrelevant operator $\Delta$ and the 
fixed point mobility $\mu^*$ using both methods.  

From a conformal field theory 
calculation, the scaling dimension of the 
leading irrelevant operator at the Kondo fixed 
point is given by \cite{ludwig91}
\begin{equation}
\Delta = 1 + \frac{2}{N+2}. \label{eq:DeltaKondo}
\end{equation}
We may also compute the same quantity using 
an $\epsilon$-expansion technique described 
in Appendix C.  By substituting $N=D+1$ 
in Eq.\ (\ref{eq:deltaDHoneycomb}), we get
\begin{equation}
\Delta = 1 + \frac{2}{N} - \frac{4}{N^2} + {\mathcal O}(\frac{1}{N^3}).
\end{equation}
This obviously agrees with 
Eq.\ (\ref{eq:DeltaKondo}) for 
large $N$.  

The next quantity we compute is the universal 
mobility $\mu^*$ of the QBM model at 
the intermediate fixed point.  
Firstly, we use the conformal 
field theory results of the Kondo model.  
Let us define the spin currents
\begin{eqnarray}
J_a & = & v_F s^z_a \label{eq:Ja} \\
& = & \frac{v_F}{4\pi} \partial_x \phi_a
 = \frac{1}{4\pi} \partial_t \phi_a,
\end{eqnarray}
and their linear combinations
\begin{eqnarray}
J^{\text{s}} & = & \sum_a \frac{1}{\sqrt{N}} J_a
 = \frac{1}{4\pi} \partial_t \phi^{\text{s}}, \\
J^{\text{sf}}_i & = & \sum_a {\mathsf O}_{ia} J_a
 = \frac{1}{4\pi} \partial_t \phi^{\text{sf}}_i. \label{eq:JsfJ}
\end{eqnarray}
Note that 
$J^{\text{s}}$ is the total spin current 
and $J^{\text{sf}}_i$ are the spin flavor 
currents.  Since the analogue of the Brownian 
particle position $r_i$ is the total spin flavor 
in each channel $S^{\text{sf}}_i=\sum_a {\mathsf O}_{ia}S^z_a$, 
the velocity $\partial_t r_i$ is mapped to the rate 
in which spin flavor is ``injected'' into channel $i$.  
Since the injection occurs at $x=0$ through the exchange 
coupling $H_J$, the spin flavor injection 
rate $\partial_tS^{\text{sf}}_i$ may be measured via the 
difference in $J^{\text{sf}}_i$ right before 
and after $x=0$.  In other words, we have a ono-to-one 
correspondence
\begin{equation}
J^{\text{sf}}_i(x=0^+) - J^{\text{sf}}_i(x=0^-)\ \leftrightarrow\ \partial_t r_i.
\end{equation}
From the above mapping and Eq.\ (\ref{eq:mur}), 
we get
\begin{equation}
\mu = \lim_{\omega\to 0} \frac{1}{2\pi|\omega|}
 \int d\tau\: (1-e^{i\omega\tau}) \Bigl< \text{T}_\tau
 \left[ J^{\text{sf}}_i(0^+,\tau) -
 J^{\text{sf}}_i(0^-,\tau) \right]
 \left[ J^{\text{sf}}_i(0^+,0) -
 J^{\text{sf}}_i(0^-,0) \right] \Bigr>_0,
\label{eq:muJ}
\end{equation}
where $\text{T}_\tau$ is the time ordering operator 
and the average $\langle\cdots\rangle_0$ is calculated with 
respect to the free non-interacting 
Hamiltonian $H_0$.  The above fixed point correlation 
functions have been exactly calculated by 
Ludwig and Affleck \cite{ludwig94,ludwig93} 
using conformal field theory.  Specifically, 
they are given by
\begin{eqnarray}
\left< \text{T}_\tau J^{\text{sf}}_i(0^+,\tau)
 J^{\text{sf}}_i(0^+,0) \right>_0
& = & \left< \text{T}_\tau J^{\text{sf}}_i(0^-,\tau)
 J^{\text{sf}}_i(0^-,0) \right>_0 = \frac{1}{2\tau^2}, \\
\left< \text{T}_\tau J^{\text{sf}}_i(0^+,\tau)
 J^{\text{sf}}_i(0^-,0) \right>_0
& = & \left< \text{T}_\tau J^{\text{sf}}_i(0^-,\tau)
 J^{\text{sf}}_i(0^+,0) \right>_0 = \frac{a^K_J}{2\tau^2},
\end{eqnarray}
where
\begin{equation}
a^K_J = \frac{S^1_{1/2}/S^1_0}{S^0_{1/2}/S^0_0}
\end{equation}
and the ``modular S-matrix'' is given by
\begin{equation}
S^j_{j'} = \sqrt\frac{2}{N+2} \sin\left[\frac{\pi(2j+1)(2j'+1)}{N+2}\right].
\end{equation}
Thus, Eq.\ (\ref{eq:muJ}) becomes
\begin{eqnarray}
\mu^* & = & \frac{1-a^K_J}{2} \\
& = & 2 \sin^2 \frac{\pi}{N+2}. \label{eq:muN}
\end{eqnarray}
This result agrees with Emery and Kivelson \cite{emery92}
for $N=2$, where the it takes the perfect 
value $\mu^*=1$ [See also Sec.\ \ref{sec:twoLead}].  

The perturbation calculation of $\mu^*$ 
in the large $D$ limit of the QBM model 
is described in Appendix C.  The result 
is given by
\begin{equation}
\mu^* = 2\pi^2 \left( \frac{1}{N^2} - \frac{4}{N^3} \right)
 + {\mathcal O}\left(\frac{1}{N^4}\right).
\end{equation}
It agrees with Eq.\ (\ref{eq:muN}) 
for large $N$.

\section{On-resonance conductance and the universal scaling}
\label{sec:spin1/2}

In this section, we will compute the on-resonance 
conductance of the two-lead spin-1/2 quantum 
dot model \cite{furusaki95} using the mapping of 
the quantum dot model to both the Kondo problem 
and the QBM model.  For convenience, 
we will actually consider a four-lead spinless model 
instead of the spin-1/2 model, since they 
are equivalent (see Sec.\ \ref{sec:overview}).  

First of all, we define the conductance matrix 
for an $N$-lead model
\begin{eqnarray}
G_{ab} = \lim_{V_1,V_2,\cdots V_N\to 0} \frac{\partial}{\partial V_b}\langle -e \partial_tQ_a \rangle.
\end{eqnarray}
It characterizes the linear current 
flowing {\em into} the $a$-th lead in response to an 
infinitesimal voltage $V_b$ in the $b$-th lead.  
The mapping described in the previous sections 
dictates that the current and the 
voltage are mapped to the spin current and the 
local channel magnetic field in a Kondo model, 
respectively.  Thus, we may alternatively compute 
the ``spin conductance'' of the Kondo model.  
Explicitly, it is defined as 
\begin{equation}
G_{ab} = -\frac{e^2}{h}\lim_{\omega\to 0} \frac{1}{2\pi|\omega|}
 \int d\tau\: (1-e^{i\omega\tau}) \Bigl< \text{T}_\tau
 \left[ J_a(0^+,\tau) - J_a(0^-,\tau) \right]
 \left[ J_b(0^+,0) - J_b(0^-,0) \right] \Bigr>,
\end{equation}
where the spin current in each Kondo 
channel $J_a$ is defined in Eq.\ (\ref{eq:Ja}).  
Using Eqs.\ (\ref{eq:JsfJ}) and (\ref{eq:muJ}), the 
above equation may be evaluated to yield 
\begin{equation}
G_{ab} = \frac{e^2}{h} \left( \frac{1}{N} - \delta_{ab} \right)\mu, \label{eq:Gab}
\end{equation}
where $\mu$ is the mobility of the corresponding 
QBM model.  

A four-lead model is equivalent to the two-lead 
spin-1/2 model if we map the lead 1 and 2 
(3 and 4) to the two spin 
channels of the source (drain) lead.  If 
we apply an infinitesimal voltage $V$ to 
the source relative to the drain, i.e., 
\begin{equation}
V_a = \left\{ \begin{array}{ll}
 V & \text{if\ } a=1, 2, \\
 0 & \text{if\ } a=3, 4, \\
\end{array} \right.
\end{equation}
then the current is given by
\begin{eqnarray}
I & = & \sum_{a=3}^4 \left< -e\partial_t Q_a \right> \\
& = & \sum_{a=3}^4 \sum_{b=1}^4 G_{ab}V_b \\
& = & \mu \frac{e^2}{h} V.
\end{eqnarray}
Substituting the on-resonance mobility $\mu^*=1/2$ 
in Eq.\ (\ref{eq:muN}) to the above equation, we get
\begin{equation}
G^* = \mu^*\frac{e^2}{h} = \frac{e^2}{2h}.
\end{equation}
Physically, this result may be understood as 
follows: On resonance at $T=0$, the conductance 
of each QPC is given by $2\mu^*e^2/h=e^2/h$, where 
there is an overall factor of two due to spin.  
Then the above $G^*$ is the series conductance 
of the two QPCs (see Sec.\ \ref{sec:overview}).  

Experimentally, $G^*$ may be measured as the peak 
value of a conductance resonance as 
one sweeps the gate voltage.  
Again, the shape of the peak follows a scaling function 
that is independent of microscopic details of the 
system.  Since the scaling dimension 
of the magnetic field $\mathcal H$ 
is given by 1/3, \cite{ludwig91} 
the resonance line shape has the form
\begin{equation}
G(\delta n_0, T) = \tilde{G}\left( \frac{\delta n_0}{T^{2/3}} \right).
\end{equation}
In principle, the explicit form of $\tilde{G}$ may be 
obtained using a Monte Carlo simulation. \cite{monteCarlo}

\section{Conclusions}
\label{sec:conc}

In this paper, we have used a QBM model 
with periodic potential to study a few different 
fundamental problems in condensed matter physics.  
Generically, by integrating out unimportant degrees 
of freedom in a path integral representation, one 
can obtain an alternative description 
of a problem in terms of a first-quantized 
quantum mechanical model of a single Brownian particle 
moving in a periodic potential.  
The remaining degrees of freedom 
are mapped to the spatial coordinates 
of the Brownian particle, whereas the integrated 
ones serve as the source of dissipation.  

We have applied this method to such 
problems as resonant tunneling 
in quantum dot and Luttinger liquid 
systems as well as multichannel Kondo problems.  
For an $N$-lead quantum dot model in the 
limit $\Delta E\ll T\ll e^2/C$, we have shown that 
the on-resonance problem is mapped 
to an ($N-1$)-dimensional 
honeycomb lattice and the off-resonance 
problem to an ($N-1$)-dimensional 
triangular lattice.  Each vector component 
of the Brownian particle position is 
the total charge in each lead, and 
the periodic potential comes from 
backscattering.  
Analogously, by mapping the total spin 
in each channel to each vector 
component of the position, we 
could derive similar mappings for 
the $N$-channel Kondo problem 
in the Toulouse limit.  

The lattice symmetry of the periodic potential 
determines the zero-temperature phase diagram.  
For symmorphic lattices, there are two 
trivial zero-temperature phases: 
If the dissipation is strong or the lattice 
sites are far apart, the Brownian particle gets 
localized and the mobility $\mu$ vanishes.  
If the dissipation is weak or the lattices sites 
are close to one another, the particle diffuses 
freely and the mobility takes its perfect maximum 
value $\mu=1$.  We have analyzed the 
phase diagrams in perturbatively accessible 
regimes.  

For non-symmorphic lattices, the extra symmetry 
between sublattices may give rise to an intermediate 
phase.  The mobility in this phase is given 
a universal value in the 
range $0<\mu<1$ and independent 
of the microscopic details of the problem.  
The intermediate phase is governed by the multichannel 
Kondo fixed point.  We have explicitly 
demonstrated that the mappings are consistent 
by computing the scaling dimension 
of the leading irrelevant operator and the fixed point 
mobility in both the QBM model and the Kondo model, 
in the large $N$ limit.  

Exploiting the mapping, we have used spin current correlation 
functions calculated by a conformal field theory 
to compute the mobility at the intermediate fixed 
point exactly.  Using the result 
\begin{equation}
\mu^* = 2 \sin^2 \frac{\pi}{N+2},
\end{equation}
we could also calculate the on-resonance 
linear conductance $G_{ab}$ between each individual lead 
for the $N$-lead quantum dot model.  
Particularly for the conventional two-lead spin-1/2 model, 
the on-resonance conductance is given by $G=e^2/2h$.  
Experimentally, this conductance may be measured from 
the peak value of the conductance resonance 
as a function of the 
gate voltage.  From the QBM theory, the resonance line 
shape is predicted to be described by a scaling function 
of the form
\begin{equation}
G(\delta n_0, T) = \tilde G \left( \frac{\delta n_0}{T^\alpha} \right).
\end{equation}
where $\alpha$ is determined from the 
scaling dimension of the local magnetic field 
at the magnetic impurity.  Specifically, $\alpha=2/3$ 
for the two-lead spin-1/2 quantum dot 
model and $\alpha=1/2$ for the spinless model.

\section*{Acknowledgments}

It is a pleasure to thank C.\ L.\ Kane for 
invaluable discussions and comments.  
K.\ A.\ Matveev, J.\ J.\ Palacios, and 
T.\ S.\ Kim are also acknowledged for 
useful discussions.  This work has been 
supported by NSF under grant DMR-98-70681.

\appendix

\section{Renormalization group analysis of the QBM model}
\label{app:RGQBM}
In this appendix, we develop a renormalization 
group analysis method for the QBM model.  
The renormalization group flow equation 
for the triangular lattice will be explicitly 
calculated as an example.  More examples can be found 
in Appendix C for $D$-dimensional honeycomb 
lattices in the large $D$ limit.  

Let us consider the small $v$ limit 
of the action as in 
Eqs.\ (\ref{eq:S0v}) through (\ref{eq:Sv}).  
It should be straightforward to perform 
a similar analysis for the small $t$ limit 
using a dual theory in Eqs.\ (\ref{eq:S0t}) 
through (\ref{eq:St}).  
The partition function is given by
\begin{eqnarray}
Z & = & \int {\mathcal D}[{\mathbf r}(\tau)]\: e^{-S_0[{\mathbf r}]-S_v[{\mathbf r}]} \\
& = & Z_0 \sum_n \frac{1}{n!} \left< (-S_v)^n \right>. \label{eq:ZSv}
\end{eqnarray}
where $Z_0=\int {\mathcal D}[{\mathbf r}]\:e^{-S_0}$ is the
the partition function for $v=0$.  
The brackets denote an average with respect to the 
free action $S_0$ and are defined as
\begin{equation}
\left< O \right> \equiv \frac{1}{Z_0} \int {\mathcal D}[{\mathbf r}]\:O e^{-S_0}.
\end{equation}
Each term in Eq.\ (\ref{eq:ZSv}) may be 
explicitly written as
\begin{equation}
\frac{1}{n!} \langle (-S_v)^n \rangle
 = \sum_{\{{\mathbf G}\}} \int_{\tau_1<\tau_2<\cdots<\tau_n}
 \left( \prod_i \frac{d\tau_i}{\tau_c}\: v_i \right)
 \left< e^{i2\pi \sum_i {\mathbf G}_i \cdot {\mathbf r}(\tau_i)} \right>,
 \label{eq:Svn}
\end{equation}
where $v_i$ is a shorthand notation 
for $v_{{\mathbf G}_i}$.  The prefactor $1/n!$ 
has been absorbed in the explicit 
time ordering.  For a non-symmorphic lattice, 
time ordering is important 
because the Brownian particle has to move back and 
forth between different sublattices.  An 
explicit calculation at $T=0$ shows that
\begin{equation}
\left< e^{i2\pi \sum_i {\mathbf G}_i \cdot {\mathbf r}(\tau_i)} \right>
 = \left\{ \begin{array}{ll}
  \prod_{i<j} \left[ \frac{\tau_c^2}{(\tau_j-\tau_i)^2+\tau_c^2}
  \right]^{-{\mathbf G}_i\cdot {\mathbf G}_j} & \text{if\ } \sum {\mathbf G}_i = 0, \\
  0 & \text{otherwise}. \\
\end{array} \right. \label{eq:correlation}
\end{equation}

Any low energy 
properties calculated from the above 
partition function must not depend on the 
cutoff $\tau_c$.  
Therefore, by renormalizing $\tau_c$, we may obtain 
a set of scaling laws that connect a given problem 
to another one with renormalized parameters.  
In the new theory with cutoff $\tau_c'=\tau_ce^\ell$, we may define the 
renormalized amplitude $v_i^R$ in such a way 
that Eq.\ (\ref{eq:ZSv}) retains the same form when 
written in terms of $v_i^R$ and $\tau_c'$.  
With the help of Eq.\ (\ref{eq:correlation}), this 
condition may be written as 
\begin{equation}
\left( \frac{v_i^R}{\tau_c e^\ell} \right)^2
 \left[ \frac{(\tau_ce^\ell)^2}{(\tau_2-\tau_1)^2+(\tau_ce^\ell)^2} \right]^{|{\mathbf G}_i|^2}
 = \left( \frac{v_i}{\tau_c} \right)^2
 \left[ \frac{\tau_c^2}{(\tau_2-\tau_1)^2+\tau_c^2} \right]^{|{\mathbf G}_i|^2},
\end{equation}
to the second order in $v$.
Assuming $\tau_c\ll\tau_ce^\ell\ll\tau_2-\tau_1$,
we get a simple scaling equation
\begin{equation}
v_i^R e^{(|{\mathbf G}_i|^2-1)\ell} = v_i. \label{eq:scaling}
\end{equation}
Differentiating both sides of Eq.\ (\ref{eq:scaling}) 
with respect to $\ell$, 
we obtain the renormalization group flow 
equation (\ref{eq:dvdl}).  

Higher order corrections may be calculated
in a similar way.  Let us consider 
any arbitrary pair of consecutive operators 
at time $\tau_i$ and $\tau_{i+1}$ in Eq.\ (\ref{eq:Svn}).
If $\tau_c<\tau_{i+1}-\tau_i<\tau_ce^\ell$, they are no longer 
distinguished in the new theory with cutoff $\tau_ce^\ell$.  
Instead, they must be treated as a 
compound operator at a single moment:
\begin{equation}
e^{i2\pi {\mathbf G}_i\cdot{\mathbf r}(\tau_i)}
 e^{i2\pi {\mathbf G}_{i+1}\cdot{\mathbf r}(\tau_{i+1})}
 \to e^{i2\pi ({\mathbf G}_i+{\mathbf G}_{i+1})\cdot{\mathbf r}(\tau_{i})}.
\end{equation}
This gives rise to the second order correction 
to Eq.\ (\ref{eq:scaling}).  This process may be 
thought of as ``decimation'' of closely 
placed operator sequences in time.  
In general, we get the $n$-th order correction by 
lumping together $n$ operators that lie 
within $\tau_ce^\ell$.

In order to show details of the
decimation procedure, we will consider 
a triangular lattice as a specific example.  
The shortest reciprocal lattice vectors ${\mathbf G}_i$ 
and $-{\mathbf G}_i$ ($i=1,2,3$) are defined 
as in Fig.\ \ref{fig:tightbinding}(a).  
We need to find all $\mathbf G$-sequences 
that add up to a single reciprocal lattice vector.  
For example, since ${\mathbf G}_1+{\mathbf G}_2+{\mathbf G}_3=0$, 
there are two sequences that are second order in $v$:
\begin{equation}
\{-{\mathbf G}_2,-{\mathbf G}_3\}, \ \{-{\mathbf G}_3,-{\mathbf G}_2\},
\label{eq:GSequence}
\end{equation}
that add up to ${\mathbf G}_1$.  When decimated, 
they give rise to a correction to the righthand 
side of Eq.\ (\ref{eq:scaling}) 
\begin{equation}
( \zeta_{\{\overline{2}\overline{3}\}} + \zeta_{\{\overline{3}\overline{2}\}} ) v^2,
\end{equation}
where
\begin{eqnarray}
& & \zeta_{\{\overline{2}\overline{3}\}}
 = \int_0^{\tau_ce^\ell} \frac{d\tau'}{\tau_c}
 \left< e^{i2\pi\left[ -{\mathbf G}_2\cdot{\mathbf r}(\tau_1)
 - {\mathbf G}_3\cdot{\mathbf r}(\tau_1+\tau')-{\mathbf G}_1\cdot{\mathbf r}(\tau_2)
 \right]} \right>_{\tau_c}
 \left/ \left(\frac{\tau_c}{\tau_2-\tau_1}\right)^{2|{\mathbf G}_1|^2}
 \right. , \\
& & \zeta_{\{\overline{3}\overline{2}\}}
 = \int_0^{\tau_ce^\ell} \frac{d\tau'}{\tau_c}
 \left< e^{i2\pi\left[ -{\mathbf G}_3\cdot{\mathbf r}(\tau_1)
 - {\mathbf G}_2\cdot{\mathbf r}(\tau_1+\tau')-{\mathbf G}_1\cdot{\mathbf r}(\tau_2)
 \right]} \right>_{\tau_c}
 \left/ \left(\frac{\tau_c}{\tau_2-\tau_1}\right)^{2|{\mathbf G}_1|^2}
 \right. .
\end{eqnarray}
Each of the above terms comes from the 
corresponding sequence in Eq.\ (\ref{eq:GSequence}).  
We have assumed $v_1=v_2=v_3=v$.  
The overlines in $\overline{2}$ 
and $\overline{3}$ denote that 
the corresponding vectors have a minus sign.  
In general, it is not easy to compute $\zeta$, but 
if $|{\mathbf G}_i|^2\approx 1$, we get
\begin{equation}
\zeta_{\{\overline{2}\overline{3}\}} = \zeta_{\{\overline{3}\overline{2}\}} = \ell.
\end{equation}
Thus the scaling equation becomes
\begin{equation}
v^R e^{(|{\mathbf G}_i|^2-1)\ell} = v + 2\ell v^2. \label{eq:scalingTri}
\end{equation}
By differentiating both sides of Eq.\ (\ref{eq:scalingTri})
with respect to $\ell$ and writing the final 
equation in terms of the renormalized
parameter $v^R$, we obtain the renormalization group flow equation.
For $|{\mathbf G}_i|^2=1+\epsilon$, it becomes 
Eq.\ (\ref{eq:dvdlTri}).

This analysis may be straightforwardly generalized to 
even higher order corrections and nonidentical $v$'s.  
In general, for every 
sequence $\{{\mathbf G}_{i_1},\cdots,{\mathbf G}_{i_n}\}$ 
such that $\sum_p {\mathbf G}_{i_p}={\mathbf G}_1$, 
there is an $n$-th order contribution 
\begin{equation}
\zeta_{\{i_1i_2\cdots i_n\}} v_{i_1}v_{i_2}\cdots v_{i_n} \label{eq:zetav}
\end{equation}
to the righthand side of Eqs.\ (\ref{eq:scaling}) 
and (\ref{eq:scalingTri}), where
\begin{equation}
\zeta_{\{i_1i_2\cdots i_n\}} =
 \int_{0<\tau_2<\tau_3<\cdots<\tau_n<\tau_ce^\ell}
 \left( \prod_{p=2}^n \frac{d\tau_p}{\tau_c} \right)
 \prod_{1\leq p<q\leq n} \left[ \frac{\tau_c^2}{(\tau_q-\tau_p)^2+\tau_c^2}
 \right]^{-{\mathbf G}_{i_p} \cdot {\mathbf G}_{i_q}}. \label{eq:zeta}
\end{equation}
It is assumed that $\tau_1=0$ in the
above integrand.  For the two-vector sequences in 
Eq.\ (\ref{eq:GSequence}), disconnected 
terms identically vanished up to the 
second order.  In general, however, they are finite 
and have to be carefully taken care of.  
Disconnected terms can be treated basically 
in the same way as above, except for that 
they are products of one non-loop and 
one or more loops, where a 
loop is a subgroup of sequences that add 
up to zero.  For example, 
there is a fourth order disconnected 
sequence $\{{\mathbf G}_1\}\{{\mathbf G}_1,{\mathbf G}_2,{\mathbf G}_3\}$.  
The first subgroup trivially adds up 
to ${\mathbf G}_1$, but the second subgroup is a loop 
[Fig.\ \ref{fig:tightbinding}(b)].  
The contribution from this process is given by
\begin{equation}
\zeta_{\{1\}}\zeta_{\{123\}} v_1^2v_2v_3,
\end{equation}
where we define $\zeta_{\{i\}}=1$.  Higher order calculations 
for a $D$-dimensional honeycomb lattice 
in the large $D$ limit is given in 
Appendix C.

\section{Renormalization group analysis of the multichannel Kondo problem}
\label{app:RGKondo}

In this appendix, we derive the 
flow equations (\ref{eq:dJzdl}) and (\ref{eq:dJpdl}) 
of the multichannel Kondo model.  In 
the boson representation, 
the partition function may be written as
\begin{equation}
Z = \int{\mathcal D}[\phi^{\text{s}},\phi^{\text{sf}}_i] e^{-\int d\tau H'},
\end{equation}
where $H'$ is the Hamiltonian given
in Eq.\ (\ref{eq:HKondoBoson}).
For small $J_\perp$, we may write it in a 
Taylor series expansion as
\begin{eqnarray}
Z & = & Z_0\sum_n\frac{1}{n!}\left(\frac{J_\perp}{2}\right)^n
 \left<\left\{\int \frac{d\tau}{\tau_c}
 \sum_{a=1}^N \left[ S^+_{\text{imp}}
 e^{-i\left(K\phi^{\text{\tiny s}}
 + \sum_i {\mathsf O}^{-1}_{ai}\phi^{\text{\tiny sf}}_i\right)}
 + \text{c.c.} \right] \right\}^n \right> \\
& = & Z_0\sum_n \left(\frac{J_\perp}{2}\right)^n
 \int_{\tau_1<\tau_2<\cdots<\tau_n} \prod_{a=1}^N
 \left(\frac{d\tau_p}{\tau_c}\right)
 \sum_{\{a_p\}} \left< e^{-i\sum_p s_p\left(K\phi^{\text{\tiny s}}
 + \sum_i {\mathsf O}^{-1}_{a_pi}\phi^{\text{\tiny sf}}_i\right)}
 + \text{c.c.} \right>, \label{eq:ZJ}
\end{eqnarray}
where $K \equiv (1-NJ_z/2)/\sqrt{N}$ and $s_p$
alternates between $\pm 1$ 
due to $S^\pm_{\text{imp}}$.  The average $\langle\cdots\rangle$ and the
free partition function $Z_0$ are calculated
for $J_\perp=0$.  After the averages are evaluated, 
the above equation becomes
\begin{equation}
Z = Z_0\sum_n \left(\frac{J_\perp}{2}\right)^n
 \int' \prod_{a=1}^N \left(\frac{d\tau_p}{\tau_c}\right) \sum_{\{a_p\}}
 e^{-\sum_{p<q} V_{pq}}\:\delta\left( \sum_p s_p {\mathsf O}^{-1}_{a_pi} \right),
\label{eq:ZPlasma}
\end{equation}
where $\int'$ implies integration over the
range $\tau_1<\tau_2<\cdots<\tau_n$, and
\begin{equation}
V_{pq} \equiv -2 s_p s_q \left( K^2 + \sum_i {\mathsf O}^{-1}_{a_pi} {\mathsf O}_{ia_q} \right)
 \ln \frac{\tau_q-\tau_p}{\tau_c}. \label{eq:V}
\end{equation}

It is convenient to view Eq.\ (\ref{eq:ZPlasma}) 
as the partition function of a plasma in the 
$\tau$-space.  There are $N$ types of 
unit charges ${\mathsf O}^{-1}_{ai}$ ($a=1,\cdots,N$).  
Since each charge is characterized by $N-1$ vector 
components ($i=1,\cdots,N-1$), it is a ``vector plasma''.  
An arbitrary single charge in the plasma may be identified by 
the position $\tau$, the charge type $a$, and 
the sign of the charge $s$.  Then, $V_{pq}$ 
is the interaction energy 
between two charges $(\tau_p,a_p,s_p)$ and $(\tau_q,a_q,s_q)$.  

We now analyze this model using 
a real-space renormalization group 
method. \cite{anderson70}  A similar 
analysis of 1D plasma 
model has been performed for a Luttinger liquid 
resonant tunneling problem in 
Ref.\ \ref{ref:kane92a}, but it differs 
from our model in that charges in their model 
are conventional scalar objects.  
In the first step of the renormalization 
group transformation, we decimate pairs 
of closely placed charges.  Suppose there is 
a $\pm{\mathsf O}^{-1}_{ai}$ charge pair 
at $\tau'$ and $\tau'+\Delta\tau$.  If $\Delta\tau$ is small 
compared to the distance between this 
pair and other charges, the pair may be 
thought of as a dipole.  The total 
interaction of this dipole and all other 
charges is given by
\begin{equation}
V^{\text{dipole}}_a (\tau';\Delta\tau) = -2\sum_q s_p s_q
 \left(K^2+\sum_i {\mathsf O}^{-1}_{ai} {\mathsf O}_{ia_q} \right)
 \Delta\tau \partial_{\tau'} \ln \frac{|\tau'-\tau_q|}{\tau_c}.
\end{equation}
If $\tau_c< \Delta\tau <\tau_c+\delta\tau_c$, the dipole is no longer 
recognizable after increasing the cutoff 
from $\tau_c$ to $\tau_c+\delta\tau_c$.  However, it induces 
effective interaction of order $\delta\tau_c/\tau_c$ 
between other charges.  When all such 
dipoles are integrated out, the 
partition function becomes
\begin{eqnarray}
Z & = & Z_0\sum_n \left(\frac{J_\perp}{2}\right)^n
 \int' \prod_{a=1}^N \left(\frac{d\tau_p}{\tau_c}\right) {\sum_{\{a_p\}}}'
 e^{-\sum_{p<q} V_{pq}}
 \left\{1 + \left(\frac{J_\perp}{2}\right)^2
 \frac{\delta\tau_c}{\tau_c} \sum_p \int_{\tau_p}^{\tau_{p+1}} 
 \frac{d\tau'}{\tau_c} \sum_a
 \left[1-V^{\text{dipole}}_a (\tau';\tau_c) \right]\right\} \\
& = & Z_0\sum_n \left(\frac{J_\perp}{2}\right)^n
 \int' \prod_{a=1}^N
 \left(\frac{d\tau_p}{\tau_c}\right) {\sum_{\{a_p\}}}'
 e^{-\sum_{p<q} V_{pq}}
 \left\{1 + \left(\frac{J_\perp}{2}\right)^2
 \frac{\delta\tau_c}{\tau_c} N \left[ \frac{1}{T\tau_c}
 - \sum_{p<q} 8s_ps_qK^2 \ln \frac{\tau_q-\tau_p}{\tau_c}
 \right] \right\} \\
& = & Z_0 e^{N J_z^2\delta\tau_c/4T\tau_c^2}
 \sum_n \left(\frac{J_\perp}{2}\right)^n
 \int' \prod_{a=1}^N
 \left(\frac{d\tau_p}{\tau_c}\right)
 {\sum_{\{a_p\}}}'
 \exp\left[-\sum_{p<q} \left( V_{pq}
 + \frac{2NJ_z^2\delta\tau_c}{\tau_c}
 s_ps_qK^2\ln \frac{\tau_q-\tau_p}{\tau_c} \right) \right].
\end{eqnarray}
The notation $\sum'_{\{a_p\}}$ 
means that the sum is only over the vectors 
that form a loop, i.e., $\sum_p s_p {\mathsf O}^{-1}_{a_pi}=0$.  
We complete the renormalization group transformation by 
rescaling $\tau\to\tau e^\ell$ where $\ell=\delta\tau_c/\tau_c$.  Comparing the 
final expression with the original partition
function in Eq.\ (\ref{eq:ZPlasma}), we obtain the 
renormalization group flow equations
\begin{eqnarray}
\frac{dJ_z}{d\ell} & = & J_\perp^2 ( 1 - \frac{N}{2} J_z ), \\
\frac{dJ_\perp}{d\ell} & = & J_z J_\perp ( 1 - \frac{N}{4} J_z ),
 \label{eq:dJpdlApp}
\end{eqnarray}
which are exact in $J_z$ and perturbative 
in $J_\perp$.  The extra higher order term in 
Eq.\ (\ref{eq:dJpdl}) has been read off 
using the mapping between the Kondo 
model and the QBM model 
in the limit $N\gg 1$ and $|J_z-2/N|\ll 1$.

\section{Perturbative calculations}
This appendix is devoted to the 
computation of physical quantities 
at the intermediate fixed point of the $D$-dimensional 
honeycomb lattice in the large $D$ 
limit.  In the first subsection, we derive 
the renormalization group flow equation 
in the Toulouse limit 
and perturbatively compute the scaling dimension of the 
leading irrelevant operator.  In the second subsection, 
we calculate the fixed point mobility to leading 
and subleading order in $1/D$.  

\subsection{Renormalization group flow equations}
\label{app:RGPert}

We use the tunneling part of 
the action in Eq.\ (\ref{eq:StHoneycomb}) 
in the small $t$ limit.  The shortest 
direct lattice vectors of a $D$-dimensional 
honeycomb lattice are given in 
Eq.\ (\ref{eq:Ra}).  We will 
continue to use the same convention 
in which ${\mathbf R}_0\in\{{\mathbf R}_i\}$ 
($\{-{\mathbf R}_i$\}) if it originates from A sublattice 
(B sublattice).  In the Toulouse limit ($c=1$), 
they satisfy
\begin{equation}
|{\mathbf R}_0|^2 = \frac{D}{D+1}.
\end{equation}
If $D$ is large, $|{\mathbf R}_0|^2\approx 1$ and $t$ is 
only slightly relevant.  The intermediate
fixed point thus lies perturbatively close 
to the localized fixed point ($t^*\approx 0$).  The 
renormalization group analysis is 
very similar to that of the small $v$ limit 
in Appendix A.  
The leading order scaling equation may be 
simply read off from Eq.\ (\ref{eq:scaling}): 
\begin{equation}
t^R e^{-\ell/(D+1)} = t. \label{eq:scalingt1}
\end{equation}

The next lowest order term is proportional to $t^3$ 
and may be computed using a similar renormalization 
group technique as in Appendix A.  There 
are three third-order sequences 
that add up to ${\mathbf R}_1$:
\begin{eqnarray}
& & \{ {\mathbf R}_1, -{\mathbf R}_1, {\mathbf R}_1 \}, \nonumber \\
& & \{ {\mathbf R}_1, -{\mathbf R}_i, {\mathbf R}_i \}, \nonumber \\
& & \{ {\mathbf R}_i, -{\mathbf R}_i, {\mathbf R}_1 \}, \quad (i\neq 1).
\end{eqnarray}
According to Eq.\ (\ref{eq:zetav}),
their contributions to the scaling
equation are
\begin{equation}
\zeta_{\{1\overline{1}1\}} t^3,
 \ \zeta_{\{1\overline{i}i\}} t^3,
 \ \zeta_{\{i\overline{i}1\}} t^3, \quad (i\neq 1), 
\end{equation}
respectively.
There are also disconnected terms
\begin{equation}
\{ {\mathbf R}_1\} \{ {\mathbf R}_i, -{\mathbf R}_i \},
\end{equation}
and their contributions are
\begin{equation}
\zeta_{\{1\}} \zeta_{\{i\overline{i}\}}.
\end{equation}
Using Eq.\ (\ref{eq:zeta}), we may 
explicitly compute the sum of all 
third order corrections:
\begin{eqnarray}
\zeta_3 & = & \zeta_{\{1\overline{1}1\}}
 + \sum_{i=2}^{D+1} \left( \zeta_{\{1\overline{i}i\}}
 + \zeta_{\{i\overline{i}1\}} \right)
 - \sum_{i=1}^{D+1} \zeta_{\{1\}} \zeta_{\{i\overline{i}\}} \\
& = & 2\ell\int_0^1 dx\left[ D\frac{(1-x)^{2/D}-1}{x^2}
 + \frac{2}{x} \right].
\end{eqnarray}
In the large $D$ limit, 
\begin{equation}
\zeta_3 = -(D+5)\ell + {\mathcal O}(1/D). \label{eq:scalingt3}
\end{equation}

In order to obtain results up to the
subleading order in $1/D$, we need to
calculate the fifth order contribution $\zeta_5$.
A calculation analogous to the above yields
\begin{eqnarray}
\zeta_5
& = & \zeta_{\{1\overline{1}1\overline{1}1\}} \nonumber \\
& + & \sum_{i\geq 2}
 \left( \zeta_{\{1\overline{1}1\overline{i}i\}}
 + \zeta_{\{1\overline{1}i\overline{i}1\}}
 + \zeta_{\{1\overline{i}1\overline{1}i\}}
 + \zeta_{\{1\overline{i}i\overline{1}1\}}
 + \zeta_{\{i\overline{1}1\overline{i}1\}}
 + \zeta_{\{i\overline{i}1\overline{1}1\}} \right) \nonumber \\
& + & \sum_{i\geq 2}
 \left( \zeta_{\{1\overline{i}i\overline{i}i\}}
 + \zeta_{\{i\overline{i}1\overline{i}i\}}
 + \zeta_{\{i\overline{i}i\overline{i}1\}} \right) \nonumber \\
& + & \sum_{2\leq i<j}
 \left( \zeta_{\{1\overline{i}i\overline{j}j\}}
 + \zeta_{\{1\overline{i}j\overline{j}i\}}
 + \zeta_{\{i\overline{i}1\overline{j}j\}}
 + \zeta_{\{i\overline{j}1\overline{i}j\}}
 + \zeta_{\{i\overline{i}j\overline{j}1\}}
 + \zeta_{\{i\overline{j}j\overline{i}1\}} \right) \nonumber \\
& - & \left[ \zeta_{\{1\overline{1}1\}}
 + \sum_{i\geq 2} \left( \zeta_{\{1\overline{i}i\}}
 + \zeta_{\{i\overline{i}1\}} \right) \right]
 \left[ \sum_{i\geq 1} \zeta_{\{i\overline{i}\}} \right] \nonumber \\
& - & \zeta_{\{1\}}
 \left[ \sum_{i\geq 1} \zeta_{\{i\overline{i}i\overline{i}\}}
 + \sum_{1\leq i<j}
 \left( \zeta_{\{i\overline{i}j\overline{j}\}}
 + \zeta_{\{i\overline{j}j\overline{i}\}} \right) \right] \nonumber \\
& + & \zeta_{\{1\}} \left[ \sum_{i\geq 1} \zeta_{\{i\overline{i}\}} \right]^2.
\end{eqnarray}
Collecting only the leading order terms in $D$, we get
\begin{equation}
\zeta_5 = 2D^2\ell + {\mathcal O}(D). \label{eq:scalingt5}
\end{equation}
Combining Eqs.\ (\ref{eq:scalingt1}), (\ref{eq:scalingt3}),
and (\ref{eq:scalingt5}) and differentiating it 
with respect to $\ell$, we obtain the following
renormalization group flow equation:
\begin{equation}
\frac{dt^R}{d\ell} = \frac{1}{D+1} t^R - (D+5) (t^R)^3 + 2D^2 (t^R)^5.
\label{eq:dtdlDHoneycomb}
\end{equation}
From $dt/d\ell=0$, 
the intermediate fixed point is given by
\begin{equation}
t^* = \frac{1}{D} - \frac{2}{D^2} + {\mathcal O}(\frac{1}{D^3}). \label{eq:tFP}
\end{equation}
By substituting $t^R=t^*+\delta t$ in Eq.\ (\ref{eq:dtdlDHoneycomb}), 
we get
\begin{equation}
\frac{d\delta t}{d\ell} = -\frac{2}{D} + \frac{6}{D^2} + {\mathcal O}(\frac{1}{D^3}).
\end{equation}
Therefore, the scaling dimension of the leading 
irrelevant operator is 
\begin{equation}
\Delta = 1 + \frac{2}{D} - \frac{6}{D^2} + {\mathcal O}(\frac{1}{D^3}).
\label{eq:deltaDHoneycomb}
\end{equation}
Since the $D$-dimensional lattice model is mapped
onto the $(D+1)$-channel Kondo model,
we may compare Eq.\ (\ref{eq:deltaDHoneycomb}) with 
Eq.\ (\ref{eq:DeltaKondo}) by setting $N=D+1$.  Clearly, 
they agree with each other.

\subsection{Fixed point mobility}
\label{app:muPert}

Now we compute the mobility at the intermediate 
fixed point obtained above.  
We will first add a source term 
\begin{equation}
S_h = \frac{1}{2} \int d\omega\: |\omega| e^{|\omega|\tau_c}
 \left[ {\mathbf h}(\omega)\cdot{\mathbf k}(-\omega) + \text{c.c.} \right],
\end{equation}
to the total action.  
From Eq.\ (\ref{eq:muk}), the mobility 
may be expressed as
\begin{eqnarray}
\mu & = & 1 - \lim_{\omega\to 0} \frac{1}{D|\omega|}
 \sum_{i=1}^D \int d\omega'\:\omega\omega'
 \left< k_i(\omega) k_i(-\omega') \right> \\
& = & 1 - \lim_{{\mathbf h},\omega\to 0} \frac{1}{D|\omega|}
 \sum_{i=1}^D \frac{1}{Z}
 \int d\omega'\:\omega\omega' \frac{\delta^2Z}{\delta h_i(-\omega) \delta h_i(\omega')},
\label{eq:muZ}
\end{eqnarray}
where
\begin{equation}
Z = \int {\mathcal D}[{\mathbf k}(\tau)]\: e^{-S[{\mathbf k}]}
\end{equation}
is the partition function.  By 
shifting the dummy variable
\begin{equation}
{\mathbf k}(\omega) \to {\mathbf k}(\omega) - {\mathbf h}(\omega),
\end{equation}
we may write the action as
\begin{equation}
S = \frac{1}{2} \int d\omega \:|\omega| e^{|\omega|\tau_c}
 \left[ |{\mathbf k}(\omega)|^2 + |{\mathbf h}(\omega)|^2 \right]
 - t \int \frac{d\tau}{\tau_c} \sum_i
 \left[ \frac{\sigma^+}{2} e^{-i2\pi{\mathbf R}_i\cdot({\mathbf k}-{\mathbf h})}
 + \text{H.c.} \right].
\end{equation}
Now it is easy to evaluate Eq.\ (\ref{eq:muZ}):
\begin{equation}
\mu = \lim_{\omega\to 0} \frac{2\pi}{|\omega|\tau_c}
 \sum_{{\mathbf R},{\mathbf R}'} \frac{{\mathbf R}\cdot{\mathbf R}'}{D}
 \int \frac{d\tau}{\tau_c} \left(1-e^{i\omega\tau}\right)
 t^2 \left<
 e^{i2\pi[{\mathbf R}\cdot{\mathbf k}(\tau)+{\mathbf R}'\cdot{\mathbf k}(0)]} \right>
\end{equation}
where $\langle\cdots\rangle$ denotes the 
connected part of an average.
In the small $t$ limit, we may Taylor series-expand 
the above equation and write
\begin{eqnarray}
t^2 \left<
 e^{i2\pi[{\mathbf R}\cdot{\mathbf k}(\tau)+{\mathbf R}'\cdot{\mathbf k}(0)]} \right>
& = & \sum_n \frac{t^{n+2}}{n!} \left<
 e^{i2\pi[{\mathbf R}\cdot{\mathbf k}(\tau)+{\mathbf R}'\cdot{\mathbf k}(0)]}
 (-S_t)^n \right>_0 \\
& = & \sum_n \frac{t^{n+2}}{n!} \int \prod_{i=1}^n
 \left(\frac{d\tau_i}{\tau_c}\right) \sum_{\{{\mathbf R}_i\}} \left<
 e^{i2\pi\left[{\mathbf R}\cdot{\mathbf k}(\tau)+{\mathbf R}'\cdot{\mathbf k}(0)
 +\sum_i{\mathbf R}_i\cdot{\mathbf k}(\tau_i)\right]} \right>_0,
\end{eqnarray}
where $\langle\cdots\rangle_0$ is the connected
part of an average over $S_0$ [Eq.\ (\ref{eq:S0k})].  
Note that the lattice vector is 
alternately chosen from $\{{\mathbf R}_i\}$ 
and $\{-{\mathbf R}_i\}$.  
The first non-vanishing term is second order in $t$
and is given by
\begin{eqnarray}
\mu^{(2)} & = & \lim_{\omega\to 0}\frac{2\pi t^2}{|\omega|\tau_c^2}
 \sum_{\mathbf R} \frac{|{\mathbf R}|^2}{D} \int d\tau (1-e^{i\omega\tau})
 \left< e^{i2\pi {\mathbf R}\cdot[{\mathbf k}(\tau)-{\mathbf k}(0)]} \right>_0 \\
& = & \lim_{\omega\to 0}\frac{2\pi t^2}{|\omega|\tau_c^2}
 \int d\tau (1-e^{i\omega\tau})
 \left(\frac{\tau_c^2}{\tau^2+\tau_c^2}\right)^{D/(D+1)} \\
& = & 2\pi^2 t^2 (|\omega|\tau_c)^{-2/(D+1)}. \label{eq:mu2}
\end{eqnarray}
From a similar calculation, 
the next order term is given by
\begin{eqnarray}
\mu^{(4)} & = & \lim_{\omega\to 0}\frac{\pi t^4}{|\omega|\tau_c^4}
 \sum_{{\mathbf R}{\mathbf R}'{\mathbf R}_1} \frac{{\mathbf R}\cdot{\mathbf R}'}{D}
 \int d\tau d\tau_1 d\tau_2 (1-e^{i\omega\tau})
 \left< e^{i2\pi \left[ {\mathbf R}\cdot{\mathbf k}(\tau)+{\mathbf R}'\cdot{\mathbf k}(0)
 + {\mathbf R}_1\cdot{\mathbf k}(\tau_1)
 - ({\mathbf R}+{\mathbf R}'+{\mathbf R}_1)\cdot{\mathbf k}(\tau_2) \right]}
 \right>_0 \\
& = & -\lim_{\omega\to 0}\frac{4\pi Dt^4}{|\omega|}
 \int' d\tau d\tau_1 d\tau_2
 \frac{1}{\left[\left(\tau_2-\tau_1\right)^2+\tau_c^2\right]
 \Bigl[ \tau^2+\tau_c^2 \Bigr]} \label{eq:mu4Tau} \\
& = & -4\pi^2 D t^4 [ 1-\ln(|\omega|\tau_c) ], \label{eq:mu4}
\end{eqnarray}
where we kept only the highest order 
terms in $D$.  In Eq.\ (\ref{eq:mu4Tau}), $\int'$ 
means an integration over the range
\begin{equation}
\tau_1<0<\tau_2<\tau \text{\quad and\quad} 0<\tau_1<\tau<\tau_2.
\end{equation}
Combining Eqs.\ (\ref{eq:mu2}) and (\ref{eq:mu4}) and 
writing them in terms of the renormalized 
tunneling amplitude $t^R$,
the mobility is given by
\begin{equation}
\mu = 2\pi^2\left[\left(t^R\right)^2-2D\left(t^R\right)^4\right].
\end{equation}
Substituting the fixed point value $t^*$ in 
Eq.\ (\ref{eq:tFP}), the above equation becomes
\begin{equation}
\mu^* = 2\pi^2 \left( \frac{1}{D^2} - \frac{6}{D^3} \right)
 + {\mathcal O}\left(\frac{1}{D^4}\right).
 \label{eq:muNperturb}
\end{equation}
When $N=D+1$, this is consistent with the 
exact solution Eq.\ (\ref{eq:muN}).

\pagebreak

\begin{table}
\begin{tabular}{c|l|ccccc}
\hline
$D$ & lattice  \{$\mathbf R$\} & $|{\mathbf G}_0|^2$ & $dv/dl$ & $v^*$ & $\Delta$ & $\mu^*$ \\
\hline \hline
& triangular & $1+\epsilon$ & $-\epsilon v+2v^2$ & $\frac{\epsilon}{2}$ & $1-\epsilon$ & $1-\frac{3\pi^2}{2}\epsilon^2$ \\
\raisebox{2ex}[0in][0in]{2} & honeycomb & $1-\epsilon$ & $\epsilon v-2v^2$ & $\frac{\epsilon}{2}$ & $1+\epsilon$ & $1-\frac{3\pi^2}{2}\epsilon^2$ \\
\hline
& FCC & $1+\epsilon$ & $-\epsilon v+(A_3+B_3)v^3$ & $\sqrt{\frac{\epsilon}{A_3+B_3}}$ & $1-2\epsilon$ & $1-\frac{16\pi^2}{3(A_3+B_3)}\epsilon$ \\
\raisebox{2ex}[0in][0in]{3} & diamond & $1-\epsilon$ & $\epsilon v-(A_3-B_3)v^3$ & $\sqrt{\frac{\epsilon}{A_3-B_3}}$ & $1+2\epsilon$ & $1-\frac{16\pi^2}{3(A_3-B_3)}\epsilon$ \\
\hline
& generalized FCC & $1+\epsilon$ & $-\epsilon v+B_Dv^3$ & $\sqrt{\frac{\epsilon}{B_D}}$ & $1-2\epsilon$ & $1-\frac{4(D+1)\pi^2}{DB_D}\epsilon$ \\
\raisebox{2ex}[0in][0in]{$D\geq 4$} & generalized diamond & $1+\epsilon$ & $-\epsilon v+B_Dv^3$ & $\sqrt{\frac{\epsilon}{B_D}}$ & $1-2\epsilon$ & $1-\frac{4(D+1)\pi^2}{DB_D}\epsilon$ \\
\hline \hline
\multicolumn{7}{l}{\quad $A_3 = 6B(1/3,1/3) = 6\frac{\Gamma(1/3)\Gamma(1/3)}{\Gamma(2/3)} \approx 31.80$} \\
\multicolumn{7}{l}{\quad $B_D = 2D\int^1_0 dx \left[ \frac{x^{2/D}+x^{-2/D}-2}{(1-x)^2} + \left(\frac{x}{1-x}\right)^{2/D} - 1 \right]$,\ ($B_3 = 4\sqrt{3}\pi \approx 21.77$, $B_\infty = 0$)} \\
\hline
\end{tabular}
\vspace{.1in}
\caption{Results of the renormalization group analysis 
near $|{\mathbf G}_0|^2=1$ for small $v$.  We 
assume $\epsilon>0$.  $\mu^*$ is the mobility and $\Delta$ 
is the scaling dimension of the most 
relevant operator, both at the fixed point $v^*$.}
\label{tab:v}
\end{table}

\begin{table}
\begin{tabular}{c|l|ccccc}
\hline
$D$ & lattice \{$\mathbf R$\} & $|{\mathbf R}_0|^2$ & $dt/dl$ & $t^*$ & $\Delta$ & $\mu^*$ \\
\hline \hline
& triangular & $1+\epsilon$ & $-\epsilon t+2t^2$ & $\frac{\epsilon}{2}$ & $1-\epsilon$ & $\frac{3\pi^2}{2}\epsilon ^2$ \\
\raisebox{2ex}[0in][0in]{2} & honeycomb & $1-\epsilon$ & $\epsilon t-3t^3$ & $\sqrt{\frac{\epsilon}{3}}$ & $1+2\epsilon $ & $\pi^2\epsilon $ \\
\hline
& FCC & $1+\epsilon$ & $-\epsilon t+4t^2$ & $\frac{\epsilon}{4}$ & $1-\epsilon$ & $\frac{\pi^2}{2}\epsilon ^2$ \\
\raisebox{2ex}[0in][0in]{3} & diamond & $1-\epsilon$ & $\epsilon t-C_3t^3$ & $\sqrt\frac{\epsilon}{C_3}$ & $1+2\epsilon $ & $\frac{8\pi^2}{3C_3}\epsilon $ \\
\hline
& generalized FCC & $1+\epsilon$ & $-\epsilon t+2(D-1)t^2$ & $\frac{\epsilon}{2(D-1)}$ & $1-\epsilon$ & $\frac{(D+1)\pi^2}{2(D-1)^2}\epsilon ^2$ \\
\raisebox{2ex}[0in][0in]{$D\geq 4$} & generalized diamond & $1-\epsilon$ & $\epsilon t-C_Dt^3$ & $\sqrt\frac{\epsilon}{C_D}$ & $1+2\epsilon $ & $\frac{2(D+1)\pi^2}{DC_D}\epsilon $ \\
\hline \hline
\multicolumn{7}{l}{\quad $C_D = D+1 - 2\int_0^1 dx \frac{1}{x^2} \left[ D(1-x)^{2/D} - D + 2x \right]$,\ ($C_2 = 3$, $C_\infty = D+5+{\mathcal O}(D^{-1})$)} \\
\hline
\end{tabular}
\vspace{.1in}
\caption{Results of the renormalization group analysis 
near $|{\mathbf R}_0|^2=1$ for small $t$.  We 
assume $\epsilon>0$.  $\mu^*$ is the mobility and $\Delta$ 
is the scaling dimension of the most 
relevant operator, both at the fixed point $t^*$.}
\label{tab:t}
\end{table}

\begin{thebibliography}{10}
\bibitem{perfectConductance}
\label{ref:perfectConductance}
This is the series conductance of two incoherent point contacts with two spin
  channels. For each contact, $G=2e^2/h$.
\bibitem{monteCarlo}
\label{ref:monteCarlo}
A similar analysis using a Monte Carlo simulation has been done for the
  fractional quantum Hall edge transport problem in Ref.\ [\ref{ref:moon93}].
\end{thebibliography}
\end{document}